\documentclass[letterpaper,11pt]{article}
\usepackage[utf8]{inputenc}

\pdfoutput=1 
\usepackage{jheppub}
\usepackage{graphicx,verbatim}
\usepackage{blindtext}
\usepackage[T1]{fontenc}
\usepackage{epsf}
\usepackage{mathtools}
\usepackage{lmodern}
\usepackage[mathscr]{euscript}
\usepackage{ytableau}
\usepackage{youngtab}
\usepackage{booktabs}
\usepackage{enumerate}

\usepackage{amsmath}
\usepackage{amsfonts}
\usepackage{amssymb}
\usepackage{mathrsfs}
\usepackage{slashed}
\usepackage{xcolor}

\def\vc{\Yvcentermath1}

\def\kq{\mathfrak{q}}

\def\rx{\mathrm{x}}

\def\bi{\mathbf{i}}
\def\bj{\mathbf{j}}
\def\bl{\mathbf{l}}
\def\bz{\mathbf{z}}

\def\bx{\mathbf{x}}
\def\ba{\mathbf{a}}

\def\bn{\mathbf{n}}

\def\bT{\mathbf{T}}

\def\bN{\mathbf{N}}
\def\bK{\mathbf{K}}

\def\BC{\mathbb{C}}
\def\BR{\mathbb{R}}
\def\BE{\mathbb{E}}
\def\BZ{\mathbb{Z}}
\def\BB{\mathbb{B}}

\def\BQ{\mathbb{Q}}

\def\BR{\mathbb{R}}
\def\BC{\mathbb{C}}
\def\CalN{\mathcal{N}}

\def\CalR{\mathcal{R}}
\def\CalZ{\mathcal{Z}}

\def\CalW{\mathcal{W}}

\def\CalO{\mathcal{O}}

\def\CalX{\mathcal{X}}

\def\ve{{\varepsilon}}

\def\tm{\mathtt{m}}

\def\EH{\EuScript{H}}

\def\EK{\EuScript{K}}

\def\p{\partial}
\def\o{\omega}

\def\beq{\begin{equation}}
\def\eeq{\end{equation}}

\title{Defect in Gauge Theory and Quantum Hall States}
\author[]{Taro Kimura${}^{\spadesuit}$ and Norton Lee${}^{\diamondsuit}$}
\affiliation{$^{\spadesuit}$\rm Institut de Mathématiques de Bourgogne, Université Bourgogne Franche-Comté, France}
\affiliation{$^{\diamondsuit}$\rm Center for Geometry and Physics, Institute for Basic Science (IBS), Pohang 37673, Republic of Korea}

\emailAdd{taro.kimura@u-bourgogne.fr}
\emailAdd{norton.lee@ibs.re.kr}

\vspace{2cm}
\abstract{
We study the surface defect in $\CalN=2^*$ $U(N)$ gauge theory in four dimensions and its relation to quantum Hall states in two dimensions.
We first prove that the defect partition function becomes the Jack polynomial of the variables describing the brane positions by imposing the Higgsing condition and taking the bulk decoupling limit.
Further tuning the adjoint mass parameter, we may obtain various fractional quantum Hall states, including Laughlin, Moore-Read, and Read-Rezayi states, due to the admissible condition of the Jack polynomial.

\vspace{1em}
}

\begin{document}

\maketitle

\section{Introduction}

The relation of low-energy physics of supersymmetric gauge theory and integrable system has been an active research for decades \cite{Gorsky:1993pe,Gorsky:1995zq,Nekrasov:1995nq}. One of the best-known story is the Seiberg-Witten curve of the $\CalN=2$ supersymmetric gauge theories can be identified as the spectral curve of the integrable systems. 
This correspondence was later extended to the quantum level by Nekrasov and Shatashivilli in \cite{Nekrasov:2009uh, Nekrasov:2009ui}, with the gauge theories subjected to the $\Omega$-deformation. This deformation introduces two parameters $(\ve_1,\ve_2)$ associated to the rotation on the two orthogonal plane in $\BR^4=\BC^2$. The partition function $\CalZ$ and BPS observables can be computed exactly by localization technique for a variety of gauge theories \cite{Nikita:I}. In the limit $(\ve_1,\ve_2) \to (0,0)$, the classical integrable system is recovered. The Nekrasov-Shatashivilli limit (NS-limit for short) $\ve_1 \to \hbar$ and $\ve_2 \to 0$ results in an $\CalN=(2,2)$ supersymmetry being preserved in the fixed plane. One expects to get the quantum integrable system.

\subsection*{From gauge theory to integrable model} 

One is naturally to ask the question of computing the wavefunction of the integrable system. The stationary state wave function, in the context of Bethe/gauge correspondence, are the vacua of the two-dimensional $\CalN=(2,2)$ theory. In order to get the stationary wavefunction, we compute the expectation value of a special observable in the two dimensional theory - a surface defect in the four dimensional theory \cite{Nikita:V,Chen:2019vvt,Chen:2020rxu, Lee:2020hfu,Jeong:2018qpc,Jeong:2017mfh}. It turns out that induction of co-dimensional two surface defect provides a powerful tool in the study of Bethe/gauge correspondence. The parameter of the defect becomes the coordinates that the wavefunction depends on. The four dimensional theory with a co-dimensional two surface defect can be realized as a theory on an orbifold. The localization computations extend so as to compute the defect partition function and expectation value of BPS observables. 

Our scope is on the class of $qq$-characters observable in the gauge theory \cite{Nikita:I}. The main statement in \cite{Nikita:II} proves certain vanishing conditions for the expectation values of the $qq$-observables, both with or without defects. These vanishing conditions, called \emph{non-perturbative Dyson-Schwinger equations}, can be used to construct KZ-type equations \cite{Knizhnik:1984nr} satisfied by the partition function \cite{Nekrasov:2021tik, jeong2021intersecting}. In the NS-limit, the KZ-equations becomes a Schr\"{o}dinger-type equation satisfied by the partition function. 

\subsection*{Jack polynomial and quantum Hall state}

The Laughlin wavefunction has provied a key to understand the quantum Hall effect (QHE). It models the simplest abelian FQH and is the building blocks of model wavefuntion of more general states, both abelian and non-abelian such as Moore-Read and Read-Rezayi state. The wavefunctions of such models, aside from the Gaussian factor which we will drop, are conformally-invariant multivariable polynomials. 
All three of Laughlin, Moore-Read, and Read-Rezayi state wavefunctions are proven to be special cases of the Jack polynomial $J^{\frac{1}{\kappa}}_\bn$ with the Jack parameter $\kappa$ taking negative rational value~\cite{Bernevig:2007nek,Bernevig:2008rda,Bernevig:2009zz}. 

\subsection*{Summary and organization}


In this paper we will establish the relations between three objects: the surface operator in the 4-dimensional $\CalN=2^*$ theory, the Jack polynomials, and fractional quantum Hall states.
The main end-result is to realize the fractional quantum Hall states as instanton partition function of 4-dimensional $\CalN=2^*$ gauge theory with the presence of full-type surface defect in the following simultaneous limits

\begin{enumerate}[(i)]
    \item Nekrasov-Shatashivili limit $\ve_2 \to 0$, \label{cond:NS}
    \item Bulk-decoupling limit $\kq = e^{2\pi i \tau} \to 0$, \label{condi:Tri} 
    \item Higgsing the Coulomb moduli parameters $\{a_\alpha\}$ to sum of adjoint mass $m$ and $\Omega$-deformation paranmeter $\ve_1$, \label{condi:Higgs}
    \item Tuning the ratio between the adjoint mass $m$ and $\ve_1$ to control the filling factor of the quantum Hall states. \label{condi:filling}
\end{enumerate}

The paper is organized as follows: 
\begin{itemize}
    \item In section \ref{sec:N=2^*} we will review the instanton partition function of $\CalN=2^*$ and prove that in the Nekrasov-Shatashivili limit $\ve_2 \to 0$ (\ref{cond:NS}) the defect partition function is the eigenfunction of the elliptic Calogero-Moser system .
    \item In section \ref{sec:bulk-decouple}, we will show that in the trogonometric limit $\tau \to i\infty$ (\ref{condi:Tri}) the Calogero-Moser Hamiltonian becomes the Laplace-Beltrami operator after a canonical transformation. The Jack polynomials are the eigenfunction of the Laplace-Beltrami operator. 
    \item In section \ref{sec:Jack} we will review some basic property of Jack polynomials. 
    \item In section \ref{sec:Quantization} we will impose Higgsing condition (\ref{condi:Higgs}) to the $\CalN=2^*$ supersymmetric gauge theory. The Higgsing truncates the infinite summation of the instanton partition function. By using the Young Tableaux representation of the instanton configuration, we prove that the defect partition function becomes the Jack polynomial after Higgsing.
    \item In section \ref{sec:gauge to Hall} we recover both the Laughlin and Moore-Read quantum Hall states from the defect partition function with a file tuning of the adjoing mass $m$ (\ref{condi:filling}). We also discuss about the admissible condition satisfiled by the Jack polynomial. 
    \item We end this paper with discussion about potential future work in section \ref{sec:discussion}.
\end{itemize}


\section{Four Dimensional $\CalN=2^*$ Gauge Theory}\label{sec:N=2^*}

We consider $\CalN=2^*$ $U(N)$ gauge theory in four dimensions with adjoint mass $m$. The vacuum of the theory is characterized by Coulomb moduli parameters $\ba=(a_1,\dots,a_N)$ and exponentiated complex gauge coupling
\begin{align}
    \kq = e^{2\pi i \tau}, \ \tau = \frac{4\pi i}{g^2} + \frac{\vartheta}{2\pi}
\end{align}
The instanton partition function can be calculated via supersymmetric localization computation in the presence of an $\Omega$-background, whose deformation parameters are $(\ve_1,\ve_2)$.
The instanton configuration is labeled by a set of Young diagrams $\boldsymbol{\lambda}=(\lambda^{(1)},\dots,\lambda^{(N)})$, $\lambda^{(\alpha)}=(\lambda^{(\alpha)}_1,\lambda^{(\alpha) }_2,\dots)$ satisfying 
\begin{align}
    \lambda^{(\alpha)}_i \geq \lambda^{(\alpha)}_{i+1} \geq 0
\end{align}
which denotes the number of boxes on each row in the Young diagrams. We define the formal sum of the exponentials
\begin{align}
    \bN = \sum_{\alpha=1}^N e^{a_\alpha}, \quad \bK = \sum_{\alpha=1}^N \sum_{(i,j)\in \lambda^{(\alpha)}} e^{a_\alpha + (i-1) \ve_1 + (j-1) \ve_2}.
\end{align}
The pseudo-measure associated to the instanton configuration is defined using the index functor $\BE$ that converts the additive Chern class character to multiplicative class
\begin{align}
    \BE \left[ \sum_{a} \mathsf{n}_a e^{\mathsf{x}_a} \right] = 
    \begin{cases}
    \displaystyle
    \prod_{a} \mathsf{x}_a^{-\mathsf{n}_a}
    & (\text{rational}) \\[1em]
    \displaystyle
    \prod_{a} (1 - e^{-\mathsf{x}_a})^{-\mathsf{n}_a}
    & (\text{trigonometric}) \\[1em]
    \displaystyle
    \prod_{a} \theta(e^{-\mathsf{x}_a};p)^{-\mathsf{n}_a}
    & (\text{elliptic})
    \end{cases}
    \label{eq:index_func}
\end{align}
where $\mathsf{n}_a \in \BZ$ is the multiplicity of the Chern root $\mathsf{x}_a$. 
$\theta(z;p)$ is the theta function defined in \eqref{theta3}.
We remark the hierarchical structure, $\theta(e^{-\mathsf{x}};p) \xrightarrow{p \to 0} 1-e^{-\mathsf{x}} = \mathsf{x} + O(\mathsf{x}^2)$.
In this paper, we mostly apply the rational convention, which corresponds to four dimensional gauge theory.
The pseudo measure associated to the instanton configuration $\boldsymbol\lambda$ is computed by:
\begin{align}
    \CalZ[\boldsymbol\lambda] = \BE \left[ (1-e^m)(\bN\bK^*+q_1 q_2\bN^*\bK-P_1P_2\bK\bK^*) \right]. 
\end{align}
$q_i=e^{\ve_i}$ are the exponentiated $\Omega$-deformation parameters with $P_i = 1-q_i$. Given a virtual character $X = \sum_{a} \mathsf{n}_a e^{\mathsf{x}_a}$ we denote by $X^* = \sum_{a} \mathsf{n}_a e^{-\mathsf{x}_a}$ its dual virtual character. 

The supersymmetric localization equates the supersymmetric partition function of the $\Omega$-deformed $\widehat{A}_0$ $U(N)$ theory of the grand canonical ensemble 
\begin{align}\label{def:inst-part}
    \CalZ_{\rm inst} (\ba,m,\vec\ve;\kq) 
    & = \sum_{\boldsymbol\lambda} \kq^{|\boldsymbol\lambda|} \CalZ(\ba,m,\vec\ve)[\boldsymbol\lambda]
\end{align}
The pseudo-measure $\CalZ[\boldsymbol\lambda]$ can be expressed in terms of products of $\Gamma$-functions
\begin{align}\label{def:inst-part-2}
    \CalZ[\boldsymbol\lambda] = \prod_{(\alpha i)\neq(\beta j)} 
    &  \frac{\Gamma(\ve_2^{-1}(x_{\alpha i}-x_{\beta j}-\ve_1))}{\Gamma(\ve_2^{-1}(x_{\alpha i}-x_{\beta j}))} \times \frac{\Gamma(\ve_2^{-1}(\mathring{x}_{\alpha i}-\mathring{x}_{\beta j}))}{\Gamma(\ve_2^{-1}(\mathring{x}_{\alpha i}-\mathring{x}_{\beta j}-\ve_1 ))} \nonumber\\
    & \times \frac{\Gamma(\ve_2^{-1}(x_{\alpha i}-x_{\beta j}-m))}{\Gamma(\ve_2^{-1}(x_{\alpha i}-x_{\beta j}-m-\ve_1))} \times \frac{\Gamma(\ve_2^{-1}(\mathring{x}_{\alpha i}-\mathring{x}_{\beta j}-m-\ve_1))}{\Gamma(\ve_2^{-1}(\mathring{x}_{\alpha i}-\mathring{x}_{\beta j}-m ))}
    \, .
\end{align}
with definition of the following parameters
\begin{align}
    x_{\alpha i} = a_\alpha + (i-1)\ve_1 + \lambda^{(\alpha)}_i \ve_2, \quad 
    \mathring{x}_{\alpha i} = a_\alpha + (i-1)\ve_1.
\end{align}
The 1-loop contribution to the partition function can be expressed in terms of $\mathring{x}$ by
\begin{align}
    {\mathcal{Z}}_\text{1-loop} = \prod_{(\alpha i)\neq(\beta j)} \frac{\Gamma(\ve_2^{-1}(\mathring{x}_{\alpha i}-\mathring{x}_{\beta j}-\ve_1 ))}{\Gamma(\ve_2^{-1}(\mathring{x}_{\alpha i}-\mathring{x}_{\beta j}))} \frac{\Gamma(\ve_2^{-1}(\mathring{x}_{\alpha i}-\mathring{x}_{\beta j}-m ))}{\Gamma(\ve_2^{-1}(\mathring{x}_{\alpha i}-\mathring{x}_{\beta j}-m-\ve_1))}
    \, .
\end{align}
The product of the 1-loop and the instanton contribution has the $\mathring{x}$ terms completely cancelled 
\begin{align}
    \CalZ_\text{1-loop} \CalZ[\boldsymbol\lambda] = \prod_{(\alpha i)\neq(\beta j)} 
    &  \frac{\Gamma(\ve_2^{-1}(x_{\alpha i}-x_{\beta j}-\ve_1))}{\Gamma(\ve_2^{-1}(x_{\alpha i}-x_{\beta j}))} \times \frac{\Gamma(\ve_2^{-1}(x_{\alpha i}-x_{\beta j}-m))}{\Gamma(\ve_2^{-1}(x_{\alpha i}-x_{\beta j}-m-\ve_1))}.
\end{align}


\subsection{Introducing Surface defect}
Recent developments in BPS/CFT correspondence \cite{Nikita:I,Nikita:V,Jeong:2018qpc} notices differential equations of two dimensional conformal field theories, such as KZ-equation \cite{Nikita:V,jeong2021intersecting,Jeong:2020uxz} and KZB-equations can be verified by adding a regular surface defect in the supersymmetric gauge theory. These conformal equations becomes eigenvalue equations of the integrable model in the Nekrasov-Shatashivilli limit (NS-limit for short) $\ve_2 \to 0$. 
See also~\cite{Bonelli:2011fq,Bonelli:2011wx} for a relation to the ($q$-)hypergeometric function.
Moreover, the surface defect is also used to discuss non-perturbative aspects in $\mathcal{N}=2^*$ theory~\cite{Ashok:2017bld,Ashok:2017lko,Ashok:2017odt,Ashok:2018zxp}, and its relation to the isomonodromic system~\cite{Bonelli:2019boe,Bonelli:2019yjd}.

The co-dimensional two surface defect is introduced in the form of an $\BZ_l$ oribifolding acting on $\BR^4=\BC_1 \times \BC_2$ by $(\bz_1,\bz_2) \to (\bz_1,\zeta \bz_2)$ with $\zeta^l = 1$. 
The orbifold modifies the ADHM construction, generating a chainsaw quiver structure \cite{K-T}. 
Such defect is characterized by a coloring function $c:[N] \to \BZ_l$ that assigns the representation $\CalR_{c(\alpha)}$ of $\BZ_l$ to each color $\alpha = 1,\dots,N$. 
Here and below $\CalR_\omega$ denotes the one-dimensional complex irreducible representation of $\BZ_l$, where the generator $\zeta$ is represented by the multiplication of $\exp\left( \frac{2\pi i \omega}{l}\right)$ for $\omega \equiv \omega + l$. In general one can consider $\BZ_l$ orbifold of any integer $l$.
A surface defect is called \emph{full-type/regular surface defect} if $l=N$ and the coloring function $c$ bijective. 
Hereafter, we consider this case with the coloring function of the form 
\begin{align}\label{def:color}
    c(\alpha) = \alpha - 1. 
\end{align}

In the presence of surface defect, the complex instanton counting parameter $\kq$ fractionalizes to $N$ coupling $(\kq_\omega)_{\o=0}^{N-1}$
\begin{align}
    \kq = \kq_0 \kq_1 \cdots \kq_{N-1}, \quad \kq_{\omega+N} = \kq_\omega.
\end{align}
The coupling $\kq_\omega$ is assigned to the representation $\CalR_\omega$ of the quiver. We also define the fractional variables, $z_\alpha$, $\alpha=1,\dots,N$, by
\begin{align} \label{def:z}
    \kq_{\o-1} = \frac{z_{\o+1}}{z_\o}, \ z_{\o+N} = \kq z_{\o}.
\end{align}
From string theory point of view, these variables $\{z_\omega\}_{\omega = 1, \dots, N}$ are interpreted as the (exponentiated) brane positions, whereas the couplings $\{\kq_\omega\}_{\omega = 0,\dots,N-1}$ are (exponentiated) distances between the branes.

The defect instanton partition function is an integration over the $\BZ_N$-invariant fields
\begin{align}
\begin{split}
    & \CalZ_\text{defect}(\ba,m,\vec{\ve},\vec{\kq};\kq) = \sum_{\boldsymbol\lambda} \prod_{\omega} \kq_\omega ^{k_\omega} \CalZ_\text{defect}[\boldsymbol\lambda] \\
    & \CalZ_\text{defect}[\boldsymbol\lambda] =  \BE \left[ (1-e^m) (\hat{\bN}\hat{\bK}^* + \hat{q}_{1} \hat{q}_2 \hat{\bN}^* \hat{\bK} - \hat{P}_1\hat{P}_2 \hat{\bK} \hat{\bK}^* ) \right]^{\BZ_N}.
\end{split}
\end{align}
Here 
\begin{align}\label{def:k_w}
     & k_\o = \# \EK_\o, \ \EK_\o = \left \{ (\alpha,(i,j)) \mid \alpha \in [N], \ (i,j)\in \lambda^{(\alpha)}, \ \alpha+j-1 = \omega \ \text{mod } N \right\}
\end{align}
denotes the number of squares in a colored Young diagram that is in the $\CalR_\omega$ representation of $\BZ_N$ orbifold. 

For the convenience of later calculation, we scale $\ve_2 \to \frac{\ve_2}{N}$ and define the shifted moduli
\begin{align}
    a_\alpha - \frac{\ve_2}{N} c(\alpha) = \tilde{a}_{c(\alpha)}
\end{align}
which are neutral under orbifolding. All the ADHM data can now be written in terms of the shifted moduli
\begin{align}
\begin{split}
    & \hat{\bN} = \sum_{\o=0}^{N-1} \bN_\o {q}_2^{\frac{\o}{N}} \CalR_\o, \quad \bN_\omega = \sum_{c(\alpha)=\omega} e^{\tilde{a}_{\omega}}; \\
    & \hat{\bK} = \sum_{\o=0}^{N-1} \bK_{\o} {q}_2^{\frac{\o}{N}} \CalR_\o, \quad \bK_{\omega} = \sum_{\alpha} e^{\tilde{a}_\alpha} \sum_{J=0}^\infty \sum_{\underset{c(\alpha)+j-1=\omega+NJ}{{(i,j)\in \lambda^{(\alpha)}}}} q_1^iq_2^J.
\end{split}
\end{align}
and 
\begin{align}
    \hat{q}_1 = q_1 \CalR_0, \ \hat{q}_2 = q_2^{\frac{1}{N}} \CalR_1, \ \hat{P}_1 = (1-q_1) \CalR_0, \ \hat{P}_2 = \CalR_0 - q_2^{\frac{1}{N}} \CalR_1.
\end{align}

The expectation value of the defect partition function $\CalZ_\text{defect}$ in the NS-limit $\ve_2 \to 0$ has the asymptotic \cite{Nikita-Shatashvili,Nikita-Pestun-Shatashvili}
\begin{align}
    \CalZ_\text{defect} = e^{ \frac{1}{\ve_2} \CalW(\ba,m,\kq,\vec\ve) } \left( \CalZ_\text{surface} (\ba,m,\vec\ve,\vec\kq) + \CalO(\ve_2) \right)
\end{align}
with the singular part is identical to the bulk instanton partition function \eqref{def:inst-part}
\begin{align}
    \CalW = \lim_{\ve_2\to 0} \ve_2 \log \CalZ_\text{inst}.
\end{align}
The leading order contribution $\CalZ_\text{surface}$ is the \emph{surface partition function} \cite{Lee:2020hfu}, with
\begin{align}
    \CalZ_\text{surface} = \sum_{\boldsymbol\lambda} \prod_{\o=0}^{N-2} \kq_\o^{k_\o} \left[ (1-e^m) \frac{\sum_{\o=1}^{N-1} F_\o (F_{\o+1}-F_\o) }{P_1^*} \right]
\end{align}
where
\begin{align}
    F_\o = \bN_{\o-1} - P_1 \bK_{\o-1} + q_2 P_1 \bK_{N-1}, \ \o=1,\dots,N.
\end{align}

\subsection{$qq$-character and eigenvalue equation}

As we have stated previously, differential equations from conformal field theories such as KZ-equation and KZB equations can be verified with the introduction of regular surface defect. The key of these verification relies on an observable called \emph{qq-character} \cite{Nikita:I}. The fundamental $qq$-character of $\CalN=2^*$ ($\widehat{A}_0$ quiver) $U(N)$ gauge theory is given by \cite{Nikita:I,Nikita:V,Chen:2019vvt}  
\begin{align}\label{def:A0qq}
    \CalX(x)[\boldsymbol\lambda] = Y(x+\ve_+) \sum_{\mu} \kq^{|\mu|} B[\mu] \prod_{(\bi,\bj)\in \mu} \frac{Y(x+s_{\bi\bj}-m) Y(x+s_{\bi\bj}+m+\ve_+)}{Y(x+s_{\bi\bj})Y(x+s_{\bi\bj}+\ve_+)}
\end{align}
with $\ve_+ = \ve_1+\ve_2$. The definition of $Y$-function is
\begin{align}
    Y(x)[\boldsymbol\lambda] = \BE \left[ -e^x (N - P_1P_2K)^* \right].
\end{align}
$\mu$ is a single Young diagram $\mu=(\mu_1,\mu_2,\dots)$ obeying
\begin{align}
    \mu_\bi \geq \mu_{\bi+1}, \ \bi=1,2,\dots. 
\end{align}
One may realize $\mu$ as a ``dual'' instanton configuration in the eight dimensional gauge origami construction \cite{Nikita:III}. Each square in $\mu$ is labeled by
\begin{align}
    s_{\bi\bj} = (\bi-1)m - (\bj-1)(m+\ve_+).
\end{align}
Let us define 
\begin{align}
    B[\mu] = \prod_{(\bi,\bj)\in \mu} B_{12}(h_{\bi\bj}m+a_{\bi\bj}\ve_+), \quad B_{12}(x) = 1 + \frac{\ve_1\ve_2}{x(x+\ve_+)}.
\end{align}
Here $a_{\bi\bj} = \mu_{\bi} - \bj$ denotes the ``arm'' associated to a given box $(\bi,\bj)$ in the Young diagram $\mu$, the $l_{\bi,\bj} = \mu_\bj^T - \bi$ for the leg of the same box. We also define $h_{\bi\bj}=a_{\bi\bj}+l_{\bi\bj}+1$. 

The $qq$-character $\CalX(x)[\boldsymbol\lambda]$ is an Laurent polynomial in $Y(x)$ with shifted arguments defined on a specific instanton configuration $\boldsymbol\lambda$. The most important property of the $qq$-character is that its expectation value
\begin{align}
    \langle \CalX(x) \rangle = \frac{\sum_{\boldsymbol\lambda} \kq^{|\boldsymbol\lambda|} \CalZ[\boldsymbol\lambda] \CalX(x)[\boldsymbol\lambda] }{\CalZ_\text{inst}}
\end{align}
is a degree $N$ polynomial in $x$ \cite{Nikita:I}. 

In the presence of a regular surface defect, the argument $x$ is assigned to the $\CalR_\o$ representation of the orbifold and shifted to $x + \frac{\o}{N}\ve_2$. the fractional $qq$-character 
\begin{align} \label{def:qq}
    \CalX_\o (x) = \sum_{\mu} \BB^\mu_\o \prod_{(\bi,\bj)\in \mu} \frac{ Y_{\o+1-\bj}(x+s_{\bi\bj}-m) Y_{\o+1-\bj+1}(x+s_{\bi\bj}+m+\ve_+) }{ Y_{\o+1-\bj}(x+s_{\bi\bj}) Y_{\o+1-\bj+1}(x+s_{\bi\bj}+\ve_+) }
\end{align}
is build from the fractional $Y$-function
\begin{align}
\begin{split}
    & Y_\o(x) = \BE \left[ - e^x (N_\o - P_1 K_\o + P_1 K_{\o-1}) \right], \ i=1,\dots,N-1 \\
    & Y_0(x) = \BE \left[ -e^x (N_0 - P_1 K_0 + q_2 P_1 K_{N-1}) \right].
\end{split}
\end{align}
The factor $\BB^\mu_\o$ is the orbifolded version of $\kq^{|\mu|}B[\mu]$:
\begin{align}
    \BB^\mu_\o = \prod_{(\bi,\bj)\in \mu} \kq_{\o+1-\bj} \left[ 1 + \frac{\ve_1}{mh_{\bi\bj}} \right].
\end{align}
We denote the ensemble over all dual partition $\mu$ of each $\omega$ as
\begin{align}
    \BB_\o = \sum_{\mu} \BB_\o^\mu = \sum_{l_0,l_1,\dots,l_{N-1}\geq 0} \prod_{\omega'=0}^{N-1} \frac{ (l_{\o'}+\frac{\ve_1}{m})! }{(l_{\o'})! (\frac{\ve_1}{m})!} \left( \frac{z_\o}{z_{\o'}} \right)^{l_{\o'}} \kq ^l
\end{align}

The fractional $qq$-character $\CalX_\o(x)$ share the same property as the bulk $qq$-character, whose expectation value defined through
\begin{align}
    \langle \CalX_\o(x) \rangle_\text{defect} = \frac{\sum_{\boldsymbol\lambda} \CalX_\o(x)[\boldsymbol\lambda] \CalZ_\text{defect}[\boldsymbol\lambda] }{\CalZ_\text{defect}}
\end{align}
is a degree one polynomial in $x$. We expand the RHS in the large $x$ limit and denote $[ x^{-I}] \CalX_\o(x)$, $I=1,2,\dots$, as the coefficient of the $x^{-I}$ term in the Laurant expansion of $\CalX_\o(x)$.  
The following equation
\begin{align}
    \left \langle \left[ x^{-I} \right] \CalX_\o(x) \right \rangle_\text{defect} = 0, \ I = 1,2,\dots
\end{align}
can be translated to differential equations acting on the defect partition function $\CalZ_\text{defect}$. See \cite{Nikita:V} for detail. For our interest, we will look at $I=1$ case. The large $x$ expansion of $Y_\o(x)$ is
\begin{align}
    Y_\o(x) = (x-\tilde{a}_\o) \exp \left( \frac{\ve_1}{x} \nu_{\omega-1} + \frac{\ve_1\ve_2}{N x^2} k_{\o-1} + \frac{\ve_1}{x^2}(\sigma_{\o-1} - \sigma_\o) + \cdots \right)
\end{align}
where $k_\omega$ is defined in \eqref{def:k_w} and
\begin{align}
\begin{split}
    & \nu_\o = k_\o - k_{\o+1}, \ \sigma_\o = \frac{\ve_1}{2} k_\o + \sum_{(\alpha,(i,j))\in \EK_\o } \tilde{a}_\alpha + (i-1)\ve_1 + (j-1)\ve_2 .
\end{split}
\end{align}
The summation in $\EK_\o$ \eqref{def:k_w} runs through the colored squares in the Young diagram that is in the $\CalR_\o$ representaiton of the $\BZ_N$ orbifold. 
The large $x$ expansion of the fractional $qq$-character $\CalX_\omega(x)$ is equal to 
\begin{align}
\begin{split}
    \frac{1}{\BB_\o} \left[ x^{-1} \right] \CalX_{\o}(x) =
    & \frac{1}{2} (\ve_1\nu_{\o} - \tilde{a}_{\o+1} )^2 - \frac{ \tilde{a}^2_{\o+1} }{2} + \frac{\ve_2}{N} k_\o + \sigma_\o - \sigma_{\o+1}  \\
    & - m \sum_{\o'=0}^{N-1} \left[ (m+\ve_+) \nabla^\kq_{\o'} + (\ve_1\nu_{\o} - \tilde{a}_{\o+1} ) \nabla^z_{\o'} \right] \log \BB_\o
\end{split}
\end{align}
Here we define the differential operator for $\o=0,\dots,N-1$
\begin{align}
    \nabla^\kq_\o = \kq_\o \frac{\p}{\p \kq_\o}, \ \nabla^z_\o = z_\o \frac{\p }{\p z_\o} = \nabla^\kq_{\o-2}-\nabla^\kq_{\o-1}. 
\end{align}
By summing over $\omega$ and take the expectation value, we obtain a second order differential equation for the defect partition function $\CalZ_\text{defect}$
\begin{align}\label{eq:Schrodinger-1}
\begin{split}
    0 = &  \left[ \ve_1\ve_2 \kq \frac{\p}{\p \kq} + \frac{1}{2} \sum_{\omega=0}^{N-1} (\ve_1\nabla^z_{\o+2} - \tilde{a}_{\o+1})^2 + \ve_1 \frac{\ve_2}{N} \omega \nabla^z_{\o+2} - \frac{1}{2} \tilde{a}^2_{\o+1} \right. \\ 
    & \left. - m \sum_{\omega=0}^{N-1} (\nabla^z_{\o+2} \log \BB ) (\ve_1 \nabla^z_{\o+1} - \tilde{a}_{\o+1}) - m (m+\ve_+) \sum_{\o=0}^{N-1} \left( \nabla_\o^\kq \log \BB \right) \right] \CalZ_\text{defect}
\end{split}
\end{align}
where $\nabla^z_{\o+N} = \nabla^z_\o$ and
\begin{align}
    \BB = \prod_{\omega=0}^{N-1} \BB_\o = \left[ \Theta_{A_{N-1}}(\vec{z};\tau) \vec{z}^{\vec{\rho}} \kq^{-\frac{N^2}{24}} \right]^{-\frac{m+\ve_1}{m}} = \BQ^{\frac{\nu +1}{\nu}}
\end{align}
with $\nu = \frac{m}{\ve_1}$ being the ratio between the adjoint mass and $\Omega$-deformation parameter $\ve_1$. 
The function $\BQ$ is defined in \eqref{Q}. 
Here $\Theta_{A_{N-1}}(\vec{z};\tau)$ is the rank $N-1$ theta function defined as a product of Jacobi theta functions:
\begin{align}
    \Theta_{A_{N-1}}(\vec{z};\tau) = \eta(\tau)^{N-1} \prod_{\alpha>\beta} \frac{\theta_{11}(z_\alpha/z_\beta;\tau)}{\eta(\tau)}.
\end{align}
$\vec{\rho}$ is the Weyl vector of $SU(N)$ root system, whose entries are given as
\begin{equation}
    \vec{\rho}=(\rho_1,\dots,\rho_{N}); \quad \rho_\omega = \omega - \frac{N+1}{2};\quad |\vec{\rho}|^2=\sum_{\omega=1}^{N}\rho_\omega^2=\frac{N(N^2-1)}{12};\quad\vec{z}^{\vec{\rho}}=\prod_{\omega=1}^{N}z_\omega^{\rho_\omega}.
\end{equation}
See section \ref{sec:functions} for definitions of of theta function and eta function. 
By using the heat equation for $\BQ$ in \eqref{Heat eq for Q} to rewrite the $\nabla^\kq_\o$-derivative term in \eqref{eq:Schrodinger-1} to $\nabla^z_\o$-derivative. The defect partition function now obeys
\begin{align}
\begin{split}
    0 = &  \left[ \ve_1\ve_2 \kq \frac{\p}{\p \kq} + \frac{1}{2} \sum_{\omega=0}^{N-1} (\ve_1\nabla^z_{\o+2} - \tilde{a}_{\o+1})^2 + \ve_1 \frac{\ve_2}{N} \omega \nabla^z_{\o+2} - \frac{1}{2} \tilde{a}^2_{\o+1} \right. \\ 
    & \quad - \ve_1 (\nu+1) \sum_{\omega=0}^{N-1} (\nabla^z_{\o+2} \log \BQ ) (\ve_1 \nabla^z_{\o+1} - \tilde{a}_{\o+1}) \\
    & \quad \left.  - \frac{1}{2} \ve_1 (\nu+1) ((\nu+1)\ve_1+\ve_2) \left( \Delta_{\vec{z}} \log \BQ - \sum_{\o} (\nabla^z_\o \log \BQ)^2 \right) \right] \CalZ_\text{defect}
\end{split}
\end{align}

In the NS-limit $\ve_2 \to 0$, the shifted moduli approaches to the bulk moduli $\tilde{a}_\alpha \to a_\alpha$. Eq. \eqref{eq:Schrodinger-1} becomes an eigenvalue equation in the NS-limit
\begin{align}\label{eq:Schrodinger-2}
    \hat\EH \Psi = E\Psi   
\end{align}
with
\begin{align}\label{def:eigen}
    \Psi = \prod_{\o=1}^{N} z_{\omega}^{-\frac{{a}_{\o-1} }{\ve_1} + (\nu+1) \rho_\o } \CalZ_\text{defect}.
\end{align}
The Hamiltonian takes the form 
\begin{align}\label{def:EH}
    \hat\EH 
    = & \ \frac{1}{2} \sum_{\alpha=1}^N \left( \nabla^z_\alpha \right)^2 + (\nu+1) \sum_{\alpha=1}^N \left( \nabla^z_\alpha \log \Theta_{A_{N-1}} \right) \nabla^z_\alpha \nonumber \\
    & + \frac{(\nu+1)^2}{2} \sum_{\alpha=1}^N \left[ \left( (\nabla^z_\alpha)^2 \log \Theta_{A_{N-1}} \right) + (\nabla^z_\alpha \log\Theta_{A_{N-1}} )^2 \right]
\end{align}
with the eigenvalue
\begin{align}
    E = \frac{1}{\ve_1} \kq \frac{\p \CalW}{\p\kq} + \sum_{\alpha=1}^{N} \frac{{a}_{\alpha}^2}{2\ve_1^2} - \frac{(\nu+1)^2}{2} \rho_\alpha^2.
\end{align}

The differential operator on the right hand side of eigenvalue equation
\eqref{eq:Schrodinger-2} can be rewritten as the elliptic Calogero-Moser (eCM) Hamiltonian after a canonical transformation, 
\begin{align}\label{eq:CM-Hami}
    \hat{\rm H}_\text{eCM} = \sum_{\alpha=1}^N \frac{1}{2} (\nabla^z_\alpha)^2 + \nu(\nu+1) \sum_{\alpha>\beta} \wp(z_\alpha/z_\beta;\tau). \quad 
\end{align}
The complexified gauge coupling $\tau = \frac{4\pi i}{g^2}+ \frac{\vartheta}{2\pi}$ plays the role of the elliptic modulus. 
This is the Bethe/gauge correspondence between the elliptic Calogero-Moser system and four dimensional $\widehat{A}_0$ $U(N)$ supersymmetric gauge theory in the presence of regular surface defect \cite{Nikita:V,Chen:2020jla}. 
See also \cite{Negut:2009IM,Negut:2011aa} for more geometric interpretation.
The coupling constant $\nu$ is identified as the ratio between the adjoint mass of the gauge group and the $\Omega$-deformation parameter
\begin{align}
    \nu = \frac{m}{\ve_1}.
\end{align}

The parameter matching between the gauge theory and the Calogero-Moser integrable system is summmarized in the following table:
\begin{center}
    \begin{tabular}{|c|c|c|}
    \hline
    Parameter & Gauge Theory & Calogero-Moser System \\
    \hline \hline
    $z_\alpha$ & Fractional coupling & Particle coordinate \\
    \hline
    $a_\alpha$ & Moduli parameter & Rapidity 
    \\\hline 
    $m$ & Adjoing mass & Coupling constant \\
    \hline
    $\ve_1$ & $\Omega$-deformation & Planck constant \\
    \hline
\end{tabular}
\end{center}

\subsection{Bulk decoupling limit} \label{sec:bulk-decouple}

For the purpose of this paper, we will focus on the trigonometric Calogero-Moser system instead of the elliptic version. We have shown the complex gauge coupling acts as the complex modulus of the elliptic function. 
From this point of view, the bulk decoupling limit $\frac{1}{g^2} \to \infty$ ($\text{Im } \tau \to \infty$; $\kq \to 0$) corresponds to the trigonometric limit of the $\wp$-function. The elliptic Calogero-Moser Hamiltonian \eqref{eq:CM-Hami} becomes the trigonometric Calogero-Moser (tCM) Hamiltonian
\begin{align}
    \hat{\rm H}_\text{tCM} = \sum_{\alpha} \frac{1}{2} (\nabla^z_\alpha)^2 + \nu(\nu+1) \sum_{\alpha>\beta} \frac{1}{(z_\alpha-z_\beta)^2}
\end{align}

On the gauge theory side, the bulk decoupling limit $\kq \to 0$ becomes $\kq_{N-1} \to 0$ in the presence of regular surface defect. The bulk instanton, which now labeled by the $\CalR_{N-1}$ representation of the $\BZ_N$ orbifold and counted by $\kq_{N-1}$, only has the trivial (no instanton) configuration counted toward the ensemble in \eqref{def:inst-part} in the bulk. It gives a vanishing superpotential
\begin{align}
    \CalZ_\text{inst} = 1 \implies \CalW = 0.
\end{align}

Even though the bulk instanton now becomes trivial, there can be non-trivial instanton configuration on the surface. The defect partition function consists of now solely the surface defect contribution
\begin{align}
    \CalZ_\text{defect} = \CalZ_\text{surface} + \CalO(\ve_2).
\end{align}

The width of the colored partition $\lambda^{(\alpha)}$ is limited to $N - c(\alpha)$, that is
\begin{align}
\begin{split}
    & \lambda_{i}^{(\alpha)} \leq N-c(\alpha), \ \forall i=1,2,\dots \ \implies \lambda^{T,(\alpha)}_{N-c(\alpha)} = 0
\end{split} 
\end{align}
For later convenience, let us take a transpose on all Young diagrams labeling the instanton configurations, 
\begin{align}
    \lambda^{(\alpha)} \to \lambda^{T,(\alpha)}. 
\end{align}
Now the value $\lambda_{i}^{(\alpha)}$ denotes the number of squares in the colored Young diagram counted by $\kq_{c(\alpha)+i-1}$. The height is limited by
\begin{align}
    \lambda^{(\alpha)}_{N-c(\alpha)} = 0.
\end{align}

The defect Nekrasov instanton partition function consists of only the surface defect term, which is 
\begin{align}\label{eq:instdef}
    & \CalZ_\text{1-loop} \CalZ_\text{surface} \nonumber\\
    & = \sum_{\boldsymbol\lambda} \prod_{\omega=1}^{N-1} \kq_{\omega-1}^{\sum_{c(\alpha)<\omega} \lambda^{(\alpha)}_{\omega-c(\alpha)}} \BE \left[ (1-e^m) \frac{\sum_{\omega=1}^{N-1} F_\omega (F_{\omega+1}-F_{\omega})^* }{P_1^*} \right] \\
    & = \sum_{\boldsymbol\lambda} \prod_{\omega=1}^{N-1} \kq_{\omega-1}^{\sum_{c(\alpha)<\omega} \lambda^{(\alpha)}_{\omega-c(\alpha)}} 
    \prod_{c(\alpha),c(\beta)<\omega} \frac{\Gamma\left( \frac{y_{\omega,\alpha} - y_{\omega,\beta}-m}{\ve_1} \right)}{\Gamma\left( \frac{y_{\omega,\alpha} - y_{\omega,\beta}}{\ve_1} \right)} 
    \prod_{c(\alpha)<\omega+1,\ c(\beta)<\omega} \frac{ \Gamma\left( \frac{y_{\omega+1,\alpha} - y_{\omega,\beta}}{\ve_1} \right) }{ \Gamma\left( \frac{y_{\omega+1,\alpha} - y_{\omega,\beta}-m}{\ve_1} \right) } \nonumber
\end{align}
where we have defined
\begin{align}
    F_N = \sum_{\alpha=1}^N e^{{a}_\alpha}, \quad F_{\omega} = \sum_{c(\alpha)<\omega} e^{{a}_\alpha} q_1^{\lambda^{(\alpha)}_{\omega-c(\alpha)}} = \sum_{\alpha\leq\omega} e^{y_{\omega,\alpha}}, \quad y_{\omega,\alpha} := {a}_\alpha + \lambda^{(\alpha)}_{\omega-c(\alpha)} \ve_1.
\end{align}
Here we multiplied by the one loop factor $\CalZ_\text{1-loop}(\{a_\alpha\},\ve_1)$ 
\begin{align}\label{eq:1-loop}
    \CalZ_\text{1-loop} = \prod_{1\le\alpha<\beta\le N} \frac{\Gamma \left( \frac{a_\beta-a_\alpha}{\ve_1} \right)}{\Gamma\left( \frac{a_\beta-a_\alpha-m}{\ve_1} \right) }
\end{align}
to simplify the expression in \eqref{eq:instdef}.

In the bulk decoupling $\kq\to0$ limit, the theta function is reduced to the trigonometric function
\begin{align}
    \lim_{\kq\to0} \Theta_{A_{N-1}}(\vec{z}) = \prod_{\alpha>\beta}^N \frac{z_\beta-z_\alpha}{\sqrt{z_\beta z_\alpha}} = \prod_{\alpha>\beta}^N \left( {\sqrt{\frac{z_\beta}{z_\alpha}}-\sqrt{\frac{z_\alpha}{z_\beta}}} \right)
\end{align}
The second order differential operator $\hat\EH$ in \eqref{def:EH} in the decoupling limit becomes
\begin{align}
    \lim_{\kq \to 0}    \hat\EH = & \frac{1}{2}\sum_{\alpha=1}^N (\nabla^z_\alpha)^2 + \frac{\nu+1}{2} \sum_{\alpha>\beta} \frac{z_\alpha+z_\beta}{z_\alpha-z_\beta} \left( \nabla^z_\alpha - \nabla^z_\beta \right).
\end{align}
We identify $\hat\EH$ as half of the Laplace-Beltrami operator:
\begin{align}\label{def:LBop}
    \EH_{\rm LB}(\kappa) = \sum_{\alpha=1}^N \left( \nabla^z_\alpha \right)^2 + \kappa \sum_{\alpha>\beta} \frac{z_\alpha+z_\beta}{z_\alpha-z_\beta} \left( \nabla^z_\alpha - \nabla^z_\beta \right)
    \, .
\end{align}
with the following identification of the parameter
\begin{align}
    \kappa = \nu + 1
    \, .
\end{align}
We identify the defect partition function (with suitable pre-factor) as an eigenfunction of the Laplace-Beltrami operator
\begin{align}\label{eq:Schrodinger}
    \EH_{\rm LB}\Psi = E \Psi, \qquad E = \frac{\vec{{a}}^2}{\ve_1^2} - \kappa^2 \frac{N(N^2-1 )}{12}
    \, .
\end{align}


\subsection{Two particles case}

\subsubsection{Center of mass frame}

In a two body system, the center of mass frame can be separated. 
The Laplace-Beltrami operator can be rewritten in a variable $z^2=z_2/z_1$ and a center of mass variable $u^2=z_1z_2$:
\begin{align}
    \EH_{\rm LB}(\kappa) = \frac{1}{2}(\nabla^u)^2 +  \frac{1}{2}(\nabla^z)^2 + \kappa \frac{z+z^{-1}}{z-z^{-1}} \nabla^z
\end{align}
The wave function $\Psi(u,z)$ takes the separated variable form
\begin{align}
    \Psi(u,z) = u^{b} f(z)
\end{align}
with a constant $b\in \BC$. After decoupling the center of mass, we denote $z = e^{\rx}$, the Laplace-Beltrami operator becomes
\begin{align}
    \EH_{\rm LB} = \frac{1}{2}\frac{d^2}{d\rx^2} + \kappa \frac{\cosh \rx}{\sinh \rx} \frac{d}{d\rx} + \frac{b^2}{2}
\end{align}
acting on $f(\rx)$. 
We consider the following test function
\begin{align}
    f_{A,B}(\rx) = (2\sinh \rx)^A (2\cosh \rx)^B 
\end{align}
to find the eigenfunction $f(\rx)$ of the Laplace-Bletrami operator. The ansatz is chosen such that it takes the form of a polynomial in $z_1$ and $z_2$.
\begin{align}
    \EH_\text{LB} f_{A,B}(\rx) 
    = & f_{A,B}(\rx) \frac{1}{2} \left[ {A(A-1+2\kappa)}\frac{\cosh^2 \rx}{\sinh^2 \rx} + {B(B-1)}\frac{\sinh^2\rx}{\cosh^2 \rx} + (A+B)^2 + 2B \kappa +b^2\right] \nonumber\\
    = & \left[ E_{A,B} + \frac{b^2}{2} \right] f_{A,B}(\rx)
    \, .
\end{align}
In order for $f_{A,b}(\rx)$ to be a eigenfunction, we will choose $A=0, \ 1-2\kappa$ and $B=0,1$ that annihilates the $\frac{\cosh^2\rx}{\sinh^2\rx}$ and $\frac{\sinh^2\rx}{\cosh^2\rx}$ terms. 
There are four cases when $f_{A,B}(\rx)$ is an eigenfunction of $\EH_\text{LB}$:
\begin{itemize}
    \item $A=0$, $B=0$: $\Psi = f_{0,0}(\rx)=1$ is the trivial constant solution with eigenvalue $E_{0,0}=0$. 
    \item $A=0$, $B=1$: $\Psi = uf_{0,1}(\rx)=J_{(1)}(z_1,z_2) = z_1+z_2$ is Jack polynomial defined on a single partition of integer 1, with its eigenvalue $E_{0,1} = \kappa+\frac{1}{2}$. 
    \item $A=1-2\kappa$, $B=0$: $\Psi = u^{1-2\kappa} f_{1-2\kappa,0}(\rx) = (z_1-z_2)^{1-2\kappa}$ is the Laughlin state, with its eigenvalue $E_{1-2\kappa,0}=\frac{1}{2}-\kappa$.
    \item $A=1-2\kappa$, $B=1$: $f_{1-2\kappa,1}(\rx) = f_{2-2\kappa,0}'(\rx)$. Its eigenvalue is $E_{1-2\kappa,1} = 2-2\kappa$.
\end{itemize}
Here we choose $b$ such that $\Psi$ is a polynomial in $z_1$ and $z_2$

\subsubsection{Defect partition function}
The defect instanton configuration is given by a single column Young diagrams, $\lambda^{(1)} = (k)$. 
Let us recall that we have chosen the coloring function $c(\alpha)=\alpha-1$ in \eqref{def:color}.
\begin{align}
    F_1 = e^{a_1}q_1^k, \quad F_2 = e^{a_1} + e^{a_2}.
\end{align}
The defect instanton partition function \eqref{eq:instdef} takes the form 
\begin{align}\label{eq:inst-de-N=2}
    \CalZ_{\rm defect}
    & = \sum_{k=0}^\infty \kq_0^k \prod_{j=1}^k \frac{j\ve_1+m}{j\ve_1} \frac{a_1-a_2+j\ve_1+m}{a_1-a_2+j\ve_1} \nonumber\\
    & = {}_2 F_1 \left( \kappa ; \kappa + \frac{a_1-a_2}{\ve_1}; 1 + \frac{a_1-a_2}{\ve_1}; \frac{z_2}{z_1} \right)
\end{align}
where ${}_2 F_1$ is the hypergeometric function.
The eigenfunction $\Psi$ in \eqref{def:eigen} is given by
\begin{align}
    \Psi (a_0,a_1,\kappa;z_1,z_2) = \prod_{\alpha=1,2} z_\alpha^{-a_{\alpha}+\kappa \rho_\alpha} \CalZ_{\rm defect} = z_1^{-a_1-\frac{\kappa}{2}} z_2^{-a_2+\frac{\kappa}{2}}  \CalZ_{\rm defect}.
\end{align}
As we have proven before, the eigenvalue is given by \eqref{eq:Schrodinger}: 
\begin{align}
    \EH_\text{LB} \Psi = \left( \frac{a_1^2+a_2^2}{\ve_1^2} - \frac{\kappa^2}{2} \right) \Psi.
\end{align}
It is obvious that $\Psi$ is not a polynomial in $z_1$ nor $z_2$.

In the context of gauge theory, it's natural to consider the positive adjoint mass $\nu = \frac{m}{\ve_1} \geq 0$. One may also consider the limit $\nu \to 0$, which is the limit that the $\CalN=2^*$ gauge theory recovers the $\CalN=4$ symmetry. In such a case, all instanton configurations share the same pseudo-weight in the ensemble, giving
\begin{align}
\begin{split}
    & \CalZ_\text{defect} = \sum_{k=0}^\infty \kq_0^k = \frac{1}{1-\frac{z_2}{z_1}} \\
    \implies 
    & \Psi = \frac{z_1^{-\frac{a_1}{\ve_1}-\frac{1}{2}} z_2 ^{-\frac{a_2}{\ve_1}+\frac{1}{2}}}{1-\frac{z_2}{z_1}} = \frac{z_1^{-\frac{a_1}{\ve_1}} z_2^{-\frac{a_2}{\ve_1}} }{ \sqrt{\frac{z_1}{z_2}} - \sqrt{\frac{z_2}{z_1}} }.
\end{split}
\end{align}
The eigenvalue can be found by 
\begin{align}
    \EH_\text{LB} \Psi 
    & = \left[\frac{a_1^2+a_2^2}{\ve_1^2} -\frac{1}{2} \right] \Psi
    \, .
\end{align}

An interesting case we would like to investigate is the $\nu=-1$ case. By the identification of $\kappa=\nu+1=0$, The Laplace-Beltrami operator \eqref{def:LBop} is nothing but the Hamiltonian of free particles. On the gauge theory side, we notice that the only instanton configuration with non-vanishing pseudo-measure is the no-instanton configuration. 
\begin{align}
    \CalZ_\text{defect} & = \sum_{k=0}^\infty \kq_0^k \prod_{j=1}^k \frac{j-1}{j} \frac{a_1-a_2+j\ve_1-\ve_1}{a_1-a_2+j\ve_1} = 1
\end{align}
The wave function $\Psi$ is indeed of free particles 
\begin{align}
    \Psi (z_1,z_2) = z_1^{-\frac{a_1}{\ve_1}} z_2 ^{-\frac{a_2}{\ve_1}}
    \, 
\end{align}
with the eigenvalue being nothing but the kinetic energy 
\begin{align}
    E = \frac{a_1^2+a_2^2}{\ve_1^2}\, .
\end{align}

The hypergeometric function in \eqref{eq:inst-de-N=2} can be truncated when $\kappa \in \BZ_{<0}$. The defect partition function $\CalZ_{\rm defect}$ becomes a degree $-\kappa$ polynomial in $\kq_0 = \frac{z_2}{z_1}$. 


\section{Jack Polynomial}\label{sec:Jack}

In this section we collect some facts about the Jack polynomial.
See, e.g., \cite{macdonald1998symmetric,stanley1989some} for more details.
A Jack polynomial $J_{\bn}^{\frac{1}{\kappa}}(z_1,\dots,z_N)$ is a symmetric polynomial in variables $\{z_1,\dots,z_N\}$ labeled by the partition $\bn = (n_1,n_2,\dots,n_N)$:
\begin{align}
    n_i \geq n_{i+1} \geq 0, \qquad i = 1,\dots,N-1.
\end{align}
and a parameter $\kappa$. 
In the context of QHE, the partition $\bn$ can be represented as a (bosonic) occupation number configuration $\bl(\bn)=\{l_\tm(\bn),\ \tm=0,1,2,\dots\}$ of each of the lowest Landau level (LLL) orbits of angular momentum $L = \tm \hbar$, where for $\tm>0$ the number $l_\tm(\bn)$ is the multiplicity of $\tm$ in $\bn$. 
Given a partition $\bn=(n_1,n_2,\dots,n_N)$, let
\begin{align}
    M_{\bn} = \sum_{\sigma \in \mathfrak{S}_N} z_1^{\bn_{\sigma_1}} \cdots z_N^{\bn_{\sigma_N}}
\end{align}
be the orbit sum. The $\sigma \in \mathfrak{S}_N$ is a permutation of the set $\{1,\dots,N\}$.  
When $\kappa \to 0$, $J^{\frac{1}{\kappa}}_\bn \to M_\bn$ is the monomial wavefunction of the free boson state with occupancy number $\bl(\bn)$. 

The Dunkl operator $D_i$ is defined by
\begin{align}
    D_i = z_i\frac{\partial}{\partial z_i} + \frac{\kappa}{2} \sum_{j \neq i} \frac{z_i+z_j}{z_i-z_j} (1-\sigma_{ij}).
\end{align}
Here
\begin{align}
    (\sigma_{ij}f)(\cdots,x_i,\cdots,x_j,\cdots) = f(\cdots,x_j,\cdots,x_i,\cdots)
\end{align}
are the operators that exchange of the $i$-th variable and the $j$-the variable and the differentiation operations with respect to those variables. 
Two Dunkl operators commute with each other
\begin{align}
    \left[ D_i, D_j \right] = 0,
\end{align}
such that
\begin{align}
    \sum_{i=1}^N D^2_i = \sum_{i=1}^N \left( z_i \frac{\partial}{\partial z_i} \right)^2 - \kappa \sum_{i<j} \frac{z_i+z_j}{z_i-z_j} \left( z_i\frac{\partial}{\partial z_i} - z_j \frac{\partial}{\partial z_j} \right) .
\end{align}

It has been known that the eigenfunction of Laplace-Beltrami operator are the Jack polynomials 
\begin{align}
    \EH_{\rm LB}(\kappa) J_{\bn}^{\frac{1}{\kappa}}(z_1,\dots,z_N) = E(\bn) J_{\bn}^{\frac{1}{\kappa}}(z_1,\dots,z_N).
\end{align}
The energy spectrum is given by
\begin{align} \label{eq:eigenvalue-Jack}
    E(\bn) = e_{\bn} - e_\varnothing
\end{align}
where 
\begin{align}
    e_{\bn} = \sum_{\alpha=1}^N \left( n_\alpha + \kappa \left( \frac{N+1}{2} - \alpha \right) \right)^2.
\end{align}

Jack polynomial with negative rational value of $\kappa$ is used to construct wavefunction of fractional quantum Hall effect. In particular Laughlin, Moore-Read, and Read-Razayi fractional quantum Hall effect wave function can be explicitly written as single Jack symmetric polynomials, whose partitions $\bn$ obey the $(k,r)$-admissible condition~\cite{Bernevig:2007nek,Bernevig:2008rda,Bernevig:2009zz}
\begin{align}
    n_{i} - n_{i+k} \geq r , \ i=1,\dots,N-k.
\end{align}
and the coupling $\kappa = - \frac{r-1}{k+1}$ is set to negative rational number where $r-1$ and $k+1$ are co-prime. See \cite{Feigin:2002IMRN} for properties of Jack polynomial at negative rational coupling. 

\subsection{Concrete expressions}

We define the power sum polynomial,
\begin{align}
    p_n(x) = \sum_{i=1}^N z_i^n
    \, .
\end{align}
here we list a few the Jack polynomials $J_\bn^{\frac{1}{\kappa}}(z)$ given in terms of the power sum polynomial:
\begin{subequations}
\begin{align}
    J_{(1)}^{\frac{1}{\kappa}} (z) & = p_1 \\
    J_{(2)}^{\frac{1}{\kappa}} (z) & = \frac{1}{1+\kappa} p_2 + \frac{\kappa}{1+\kappa} p_1^2 \\
    J_{(1,1)}^{\frac{1}{\kappa}} (z) & = - \frac{1}{2} p_2 + \frac{1}{2} p_1^2 \\
    J_{(3)}^{\frac{1}{\kappa}} (z) & = \frac{2}{(1+\kappa)(2+\kappa)} p_3 + \frac{3 \kappa}{(1+\kappa)(2+\kappa)} p_2 p_1 + \frac{\kappa^2}{(1+\kappa)(2+\kappa)} p_1^3 \\
    J_{(2,1)}^{\frac{1}{\kappa}} (z) & = - \frac{1}{1 + 2 \kappa} p_3 + \frac{1 - \kappa}{1 + 2 \kappa} p_2 J_1 + \frac{\kappa}{1 + 2 \kappa} p_1^3 \\
    J_{(1,1,1)}^{\frac{1}{\kappa}} (z) & = \frac{1}{3} p_3 - \frac{1}{2} p_2 + \frac{1}{6} p_1^3 
\end{align}
\end{subequations}
The Jack polynomial $J_\bn^{\frac{1}{\kappa}}(z)$ can have divergent coefficients when $\kappa$ is an negative rational number.

\section{Higgsing the Coulomb Moduli Parameters} \label{sec:Quantization}

The wave function $\Psi$ \eqref{def:eigen} built from defect instanton partition function $\CalZ_\text{defect}$ was proven to be the eigenfunction of Laplace-Beltrami operator \eqref{def:LBop} in section 2. 
$\Psi$ is a function of fractional couplings $\{z_\alpha\}$, adjoint mass $m$, $\Omega$-deformation parameter $\ve_1$, and Coulomb moduli parameters $\{a_\alpha\}$. 
\begin{align}
    \Psi = \prod_{\omega=1}^{N} z_\o^{-\frac{a_{c^{-1}(\omega)}}{\ve_1} + \kappa\rho_\omega } \CalZ_{\rm defect}.
\end{align}
Here we consider a general bijective coloring function $c(\alpha): \{1,\dots,N\} \to \{0,\dots,N-1 \} $. $\Psi$ is usually an infinite series of $z_\alpha$'s. 

In this section we will demonstrate how the wave function $\Psi$ becomes a Jack polynomial: By fine-tuning the Coulomb moduli parameters $\{a_\alpha\}$ with respected to the adjoint mass $m$, the infinite instanton summation is reduced to a finite number of terms. 
This finite summation can be recast as a summation over the Young Tableaux. The summation is identifed as the combinatorial formula of the Jack polynomial \cite{knop1996recursion}.

To put the system on a circle, we denote $z_\omega = e^{2\pi i \rx_\omega}$ with a periodic variable $\rx_\omega \equiv \rx_\omega + 1$. A quantization condition for moduli parameters $a_\alpha$ shall be imposed for the wave function $\Psi$ to be single valued:
\begin{align} \label{eq:quant}
    -\frac{a_{c^{-1}(\omega)}}{\ve_1} + \kappa \rho_\omega = n_{c^{-1}(\omega)} \in \BZ.
\end{align}
This gives
\begin{align}
    { a_{c^{-1}(\omega)} - a_{c^{-1}(\omega+1)} } = -m-\ve_1 + ( n_{c^{-1}(\omega+1)} - n_{c^{-1}(\omega)} ) \ve_1 , \quad n_\alpha \in\BZ
\end{align}
The combination of $-m-\ve_1$ might seem weird at first. One can understand this by putting the gauge theory in the framework of gauge origami \cite{Nikita:III}. In the gauge origami, the adjoint mass $m$ is realized by $\Omega$-deformation parameter $\ve_3$ on the third complex plane $\BC_3$. The combination $-m-\ve_+ = \ve_4$ is the $\Omega$-deformation parameter on the fourth complex plane $\BC_4$, which becomes $-m-\ve_1$ in the $\ve_2 \to 0$ limit. 
Hence, in terms of the $\Omega$-deformation parameters, the Jack polynomial parameter is written as
\begin{align}
    \nu = \frac{\ve_3}{\ve_1}
    \, , \qquad
    \kappa = \frac{\ve_3 + \ve_1}{\ve_1} = \lim_{\ve_2 \to 0} \left( - \frac{\ve_4}{\ve_1} \right)
    \, .
    \label{nu_epsilon}
\end{align}
The condition \eqref{eq:quant} imposes a locus on the Higgs branch where it meets the Coulomb branch, known as the {\it root of Higgs branch} for $\CalN=2^*$ theory~\cite{Chen:2012we}. The physical interpretation of $n_\alpha$ is turning on a magnetic flux in the 23-direction for the $\alpha$-th $U(1)$ factor in the $U(N)$ gauge group
\begin{align}
    \frac{1}{2\pi} \int ( F_{23} )_\alpha dx_2 \wedge dx_3 = n_\alpha.
\end{align}
We denote the set of these $U(1)$ fluxes by $\mathbf{n} = (n_\alpha)_{\alpha = 1,\ldots,N}$.

\begin{figure}
    \centering
    \includegraphics[width=0.5\textwidth]{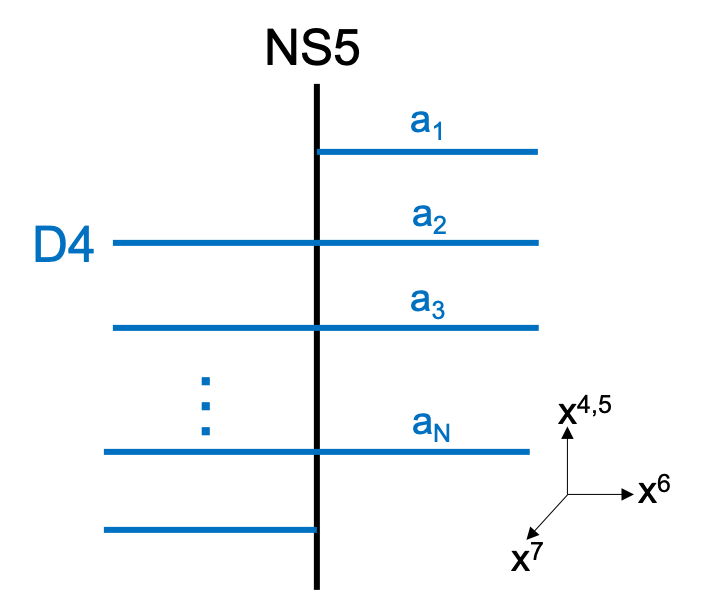}
    \caption{The Higgsing branch root.}
    \label{fig:NS5-D4}
\end{figure}

We can realize the quantization in the D-brane construction of $\CalN=2^*$ gauge theory. Let us first consider the case with the absence of magnetic flux. The mass of the adjoint hypermultiplet is realized by the $\Omega$-deformation on $\BR_{45}^2=\BC_3$ space. The two ends of the D4 brane on the NS5 no longer align by the twisted boundary condition. In particular, this allows all D4 branes to join together to from a single helical D4 or a coil wrapping along the $\bx^4$ and $\bx^6$ direction. See Figure \ref{fig:NS5-D4} for illustration. 

\begin{figure}
    \centering
    \includegraphics[width=0.5\textwidth]{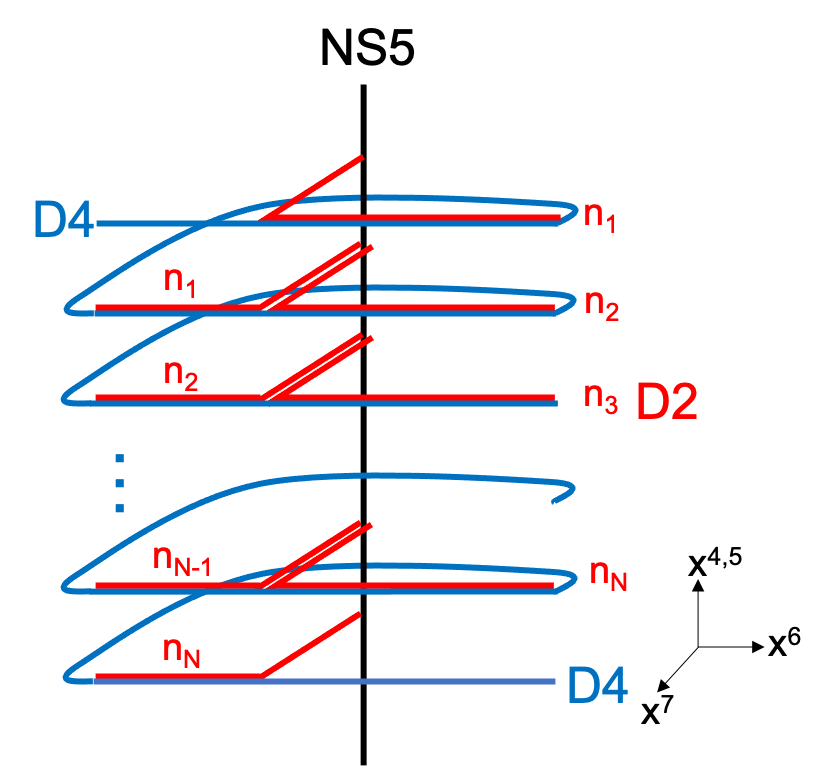}
    \caption{D-brane realization of the quantization.}
    \label{fig:NS5-D4-D2}
\end{figure}

We now turn on the magnetic flux. The quantized magnetic flux can be realized as $n_\alpha$ D2 branes "dissolving" into the $\alpha$-th D4 brane. To minimize the energy, the D2 branes prefer to stay inside the D4 brane. The compact helical structure of the D4 branes hence make the engineering of these D2 branes subtle. It is done by the following way: We take $n_\alpha$ D2 branes and stretch them from the single NS5 brane to the D4 brane in the $\bx^7$ direction, then stretch them along the $\bx^4$ and $\bx^6$ direction with a fixed $\bx^7$ inside the toroidal D4 brane, and finally stretch them back to the NS5 brane in the $\bx^7$ direction. See Figure \ref{fig:NS5-D4-D2} for the illustration. 

Near the region transverse to the D4 brane, there are $n_\alpha$ D2 branes with one orientation in the $\bx^7$ and $n_{\alpha+1}$ of D2 branes of opposite orientation. They locally annihilate each other leaving only $n_\alpha - n_{\alpha+1}$ net D2 brane stretching along the interval. The net magnetic flux from the D2 branes cancels the net magnetic charge in the D4 brane, which is $n_{\alpha} - n_{\alpha+1}$ by the opposite orientation of the adjacent D4 branes. 

\paragraph{}

We would like to argue that without loss of generality, we can consider the case 
\begin{align}
    n_1 \geq n_2 \geq \cdots \geq n_N
\end{align}
and choose the coloring function $c(\alpha) = \alpha-1$. For the generic coloring functions, the $U(1)$ fluxes $(n_{\alpha_i})_{i = 1,\ldots,N}$ can be arranged in a non-increasing order 
\begin{align}
    n_{\alpha_1} \geq n_{\alpha_2} \geq \cdots \geq n_{\alpha_N}
\end{align}
where ${\alpha_1},\dots,{\alpha_N} \in [N] = \{1,\dots,N\}$ and $\alpha_i \neq \alpha_j$ if $i\neq j$. The arrangement is nothing but a permutation of $[N]$ generated by a permutation function $s:[N] \to [N]$ such that $\alpha_i = s(i)$. We choose the coloring function $c:N \to \{0,\dots,N-1\}$ such that 
\begin{align}
    c = s^{-1} + 1. 
\end{align}
By renaming each $n_{\alpha_i}$ as $n_i$, we arrive at the case where
\begin{align}
    n_1 \geq n_2 \geq \cdots \geq n_N
\end{align}
and the coloring function $c(\alpha) = \alpha-1$. Furthermore we can set $n_N=0$ with an over all boost. The wave function $\Psi$ in \eqref{eq:instdef} can be simplified with the Higgising condition \eqref{eq:quant}: 
\begin{align}\label{eq:inst-quant}
\begin{split}
    \Psi = & \prod_{\alpha=1}^{N} z_\alpha^{n_\alpha} \sum_{\boldsymbol\lambda} \prod_{\omega=2}^{N} \kq_{\omega-1}^{\sum_{\alpha<\omega} \lambda^{(\alpha)}_{\omega-\alpha}}
    \left[ \prod_{\alpha,\beta<\omega} \prod_{j=1}^{ \lambda_{\omega-\alpha}^{(\alpha)} - \lambda_{\omega+1-\alpha}^{(\alpha)} } \frac{\kappa(\beta-\alpha)+ n_\alpha-n_\beta + \lambda_{\omega-\beta}^{(\beta)} - \lambda_{\omega-\alpha}^{(\alpha)} + j + \kappa - 1 }{\kappa (\beta-\alpha)+ n_\alpha-n_\beta + \lambda_{\omega-\beta}^{(\beta)} - \lambda_{\omega-\alpha}^{(\alpha)} + j} \right]
    \\
    & \times \prod_{\beta<\omega} \left[ \prod_{j=1}^{\lambda_{1}^{(\omega)}} \frac{\kappa (\omega-\beta) + n_\beta - n_\omega + j-1 }{\kappa (\omega-\beta) + n_\beta - n_\omega + j - \kappa } \right] 
    \left[ \prod_{j=1}^{\lambda_{\omega-\beta}^{(\beta)} } \frac{\kappa (\beta-\omega) + n_\omega - n_\beta - \lambda^{(\omega)}_1 + j -1 + \kappa }{\kappa(\beta-\omega) + n_\omega - n_\beta - \lambda^{(\omega)}_1 + j } \right].
\end{split}
\end{align}
Here we assume $n_1\geq n_2 \geq \cdots \geq n_{N} = 0$. 
In order to see $\Psi$ consists of only a finite number of terms, we notice that the instanton configuration $\boldsymbol\lambda$ that counts toward in the ensemble must obey
\begin{align}\label{eq:Young-condi}
    n_{\beta} - n_{\beta+1} + \lambda_{\omega-\beta}^{(\beta+1)} \geq \lambda_{\omega-\beta}^{(\beta)}
\end{align}
for $1 \leq \beta < \omega \leq N$ to have a non-vanishing pseudo measure.  Remember that the instanton Young diagram is limited in height $\lambda^{(\beta)}_{N+1-\beta} = 0$. By iteration, we obtain
\begin{align}\label{eq:Young-condi-2}
    n_{\alpha} - n_{\beta} + \lambda^{(\beta)}_{\omega-\alpha} \geq \lambda^{(\alpha)}_{\omega-\alpha}
\end{align}
for any $\alpha=1,\dots,\beta < \omega$.
This restricts the length of each row in Young diagram $\lambda^{(\alpha)}$. In particular when $\beta = N+1-\omega+\alpha$, we have
\begin{align}\label{eq:Young-condi-3}
    \lambda_{\omega-\alpha}^{(\alpha)} \leq n_\alpha - n_{N+1-\omega+\alpha} 
\end{align}
leaving only a finite number of terms in the ensemble in \eqref{eq:inst-quant}. 

The wave function $\Psi$ in \eqref{eq:inst-quant} has eigenvalue
\begin{align}
    \EH_\text{LB} \Psi = E(\bn) \Psi, \ E(\bn) = \sum_{\alpha=1}^{N} \left( n_\alpha - \kappa \rho_\alpha \right)^2 - \kappa^2 \frac{N(N^2-1)}{12}
\end{align}
which matches with the Jack polynomial spectrum \eqref{eq:eigenvalue-Jack}.

\paragraph{Example 1:} Let us consider the simplest case $\bn = {\tiny \yng(1)}$ with $n_1 = 1$, $n_2 = \cdots = n_{N-1} = n_N = 0$. The only instanton configuration that counts toward the ensemble is a single column Young diagram, $\lambda_i^{(\beta)} = 0$ if $\beta>1$. The instanton configuration is of single column of length one
\begin{align}
    \lambda^{(1)}_i = \begin{cases}
    1 & (i \leq l) \\
    0 & (i > l)
    \end{cases}
\end{align}
for $l = 0,1,\dots,N-2$.
We find the wave function is of the form: 
\begin{align}
    \Psi & = z_1 + \sum_{l=2}^{N} z_l \left[ \prod_{\beta=0}^{l-1}  \frac{\kappa (\beta+1)}{\kappa \beta + 1} \right] \prod_{\omega=1}^{l} \frac{\kappa \omega - \nu }{\kappa \omega } = z_1 + z_2 + \cdots + z_{N} 
\end{align}
This is the first power sum symmetric polynomial of $(z_1,\ldots,z_{N})$, which agrees with the corresponding Jack polynomial $J_{(1)}(z_1,\dots,z_{N})$.

\paragraph{Example 2:} Consider $\bn = {\tiny \yng(2)}$ with $n_1 = 2$, $n_2 = \cdots = n_{N} = 0$. The instanton partition sums over the Young diagram 
\begin{subequations}
\begin{align}
    \lambda_i^{(1)} & = \begin{cases}
    2 & (i \leq l_2) \\
    1 & (l_2 < i \leq l_1) \\
    0 & (l_1 < i \leq N-2) \\
    \end{cases}, \quad 0 \leq l_2 \leq l_1 \leq N-2; \\
    \lambda^{(\beta)}_i &= 0, \qquad \beta>1.
\end{align}
\end{subequations}
With some deliberate calculation, we find the wave function takes the following form
\begin{align}
    \Psi = \sum_{i=1}^{N} z_i^2 + \frac{2\kappa}{1+\kappa} \sum_{1 \le i<j \le N} z_i z_j = \sum_{i=1}^{N} z_i^2 + \frac{2\kappa}{1+\kappa} \sum_{1 \le i<j \le N} z_i z_j.
\end{align}
Define the power sum polynomial
\begin{align}
    p_j(z) = \sum_{i=1}^{N} z_i^j
\end{align}
such that $\Psi$ can be rewritten as
\begin{align}
    \Psi = \frac{1}{1+\kappa} p_2 + \frac{\kappa}{1+\kappa} p_1^2 = J_{(2)}^{-\frac{1}{\kappa}}(z_1,\dots,z_{N})
\end{align}
The wave function is identified as the Jack polynomial defined on partition $\bn = {\tiny \yng(2)}$.



\paragraph{Example 3:} 
The defect instanton partition function of $U(2)$ theory is an ensemble over single rowed Young diagram: 
\begin{align}
    \Psi = z_1^{n_1} \sum_{\lambda_1^{(1)}=0}^{n_1} \left( \frac{z_2}{z_1} \right)^{\lambda_1^{(1)}} \left[ \prod_{j=1}^{\lambda_1^{(1)} } \frac{j+\nu}{j} \frac{ n_1+1 - j }{n_1+\nu+1 - j} \right]
\end{align}
Here we list value of $\Psi_{n_1}$ of the first few value of $n_1$: 
\begin{itemize}
    \item $\Psi_0 = 1$. 
    \item $\Psi_1 = z_1 + z_2 = p_1$. 
    \item $\displaystyle \Psi_2 = z_1^2 + \frac{2+2\nu}{2+\nu}z_1z_2 + z_2^2 = \frac{\kappa}{1+\kappa} p_1^2 + \frac{1}{1+\kappa} p_2^2$. 
    \item $\displaystyle \Psi_3 = z_1^3 + \frac{3+3\nu}{3+\nu}z_1^2z_2 + \frac{3+3\nu}{3+\nu}z_1z_2^2 + z_2^3 = \frac{2}{(\kappa+1)(\kappa+2)} p_3 + \frac{3\kappa}{(\kappa+1)(\kappa+2)}p_2p_3 + \frac{\kappa^2}{(\kappa+1)(\kappa+2)}p_1^3$. 
    \item $\displaystyle \Psi_4 = z_1^4 + \frac{4\kappa}{\kappa+3} z_1^3 z_2 + \frac{6\kappa(\kappa+1)}{(\kappa+2)(\kappa+3)} z_1^2z_2^2 + \frac{4\kappa}{\kappa+3} z_1z_2^3 + z_2^4$.
\end{itemize}
For general $n_0$, it has poles at
\begin{align}
    \kappa = 1-n_1, \ 2-n_1, \ \dots, -\left[ \frac{n_1}{2} \right]
\end{align}

\paragraph{Example 4:} Here we consider $\bn = (n_1,n_2,n_3) = (1,1,0)$ for $N=3$ case. The instanton configuration $\boldsymbol\lambda$ must satisfy 
\begin{align}
    \lambda_1^{(1)} \leq \lambda_1^{(2)}, \ \lambda_1^{(2)} \leq 1, \ \lambda_2^{(2)} = 0.
\end{align}
to have non-vanishing contribution toward the ensemble. We obtain
\begin{align}
    \Psi_{(1,1)} & = z_1 z_2 \left[ 1 + \frac{z_3}{z_2} \frac{\kappa^2}{\kappa^2} + \frac{z_2}{z_1} \frac{z_3}{z_2} \frac{\kappa^3}{\kappa^3}  \right] = z_1 z_2 + z_1 z_3 + z_2 z_3. \nonumber\\
    & = J_{(1,1)}(z_1,z_2,z_3)
\end{align}
which agrees with the Jack polynomial defined based on the partition $\bn = {\tiny \vc\yng(1,1)}$. 

\paragraph{Example 5:} We now consider $N=3$, $\bn=(n_1,n_2,n_3) = (2,1,0)$. The instanton configuration needs to satisfy
\begin{align}
    \lambda_2^{(1)} \leq \lambda_1^{(1)} \leq \lambda_1^{(2)}+1, \ \lambda_1^{(2)} \leq 1, \ \lambda_2^{(1)} \leq 1, \ \lambda^{(3)}_1 = 0
\end{align}
to have non-vanishing pseudo-weight in the the ensemble. 
There are seven instanton configurations that meet the above requirements:
\renewcommand{\arraystretch}{1.3}
\begin{center}
    \begin{tabular}{|c|c|c|c|c|c|}
    \hline
    $\boldsymbol\lambda$ & $\lambda_1^{(1)}$ & $\lambda^{(1)}_2$ & $\lambda_1^{(2)}$ & counting & measure \\
    \hline \hline
    $({\tiny \vc \varnothing, \varnothing,\varnothing})$ & 0 & 0 & 0 & $z_1^2z_2$ & 1 \\
    \hline
    $({\tiny \vc \yng(1), \varnothing,\varnothing})$ & 1 & 0 & 0 & $z_1z_2^2$ & 1 \\
    \hline 
    $({\tiny \vc \yng(1,1), \varnothing,\varnothing})$ & 1 & 1 & 0 & $z_1z_2z_3$ & $\frac{2\kappa}{\kappa+1}\frac{\kappa+2}{2\kappa+1}$ \\
    \hline 
    $({\tiny \vc \varnothing, \yng(1),\varnothing})$ & 0 & 0 & 1 & $z_1^2z_3$ & 1 \\
    \hline
    $({\tiny \vc \yng(1), \yng(1),\varnothing})$ & 1 & 0 & 1 & $z_1z_2z_3$ & $\frac{2\kappa}{\kappa+1}$ \\
    \hline 
    $({\tiny \vc \yng(1,1), \yng(1),\varnothing})$ & 1 & 1 & 1 & $z_1z_3^2$ & 1\\
    \hline 
    $({\tiny \vc \yng(2), \yng(1),\varnothing})$ & 2 & 0 & 1 & $z_2^2z_3$ & 1\\
    \hline
    $({\tiny \vc \yng(2,1), \yng(1) ,\varnothing})$ & 2 & 1 & 1 & $z_2z_3^2$ & 1\\
    \hline
\end{tabular}
\end{center}
\renewcommand{\arraystretch}{1}
This gives
\begin{align}
    \Psi_{(2,1)} & = z_1^2 z_2 + z_1^2 z_3 + z_1 z_2^2 + z_1 z_3^2 + z_2^2 z_3 + z_2 z_3^2 + \frac{6\kappa }{2\kappa+1} z_1z_2z_3 \nonumber\\
    & = -\frac{1}{2\kappa+1} p_3 + \frac{1-\kappa}{2\kappa+1} p_2p_1 + \frac{\kappa}{2\kappa+1} p_1^3 \\
    & = J_{(2,1)}^{\frac{1}{\kappa}}(z_1,z_2,z_3) \nonumber
\end{align}
Again we see the wave function $\Psi$ is the Jack polynomial defined on the same partition $\bn = {\tiny \Yvcentermath1 \yng(2,1)}$. 

\subsection{Young Tableaux representation}

We will now introduce an alternative way to denote the instanton configuration $\boldsymbol\lambda$ that the ensemble in \eqref{eq:inst-quant} sums over. 

Let us consider a semi-standard Young Tableaux $\bT_\bn[{\boldsymbol\lambda}=\varnothing]$ of shape $\bn$. 
The initial reading of each box in the $\alpha$-th row is $\alpha$. 
We define Young Tableaux $\bT_\bn[\boldsymbol\lambda]$ based on an instanton configuration $\boldsymbol\lambda$ by the following procedure: Starting with $\bT_\bn[{\boldsymbol\lambda}=\varnothing]$, we increase the reading of the last $\lambda^{(\alpha)}_i$ squares at $\alpha$-th row (with the counting starts from the left) by one and repeat the process for $i=1,\dots,N-\alpha$. On individual row this process guarantees that the reading stays non-decreasing when moving towards right since $\lambda^{(\alpha)}_i \geq \lambda^{(\alpha)}_{i+1}$. 
For the $j$-th square in the $\alpha$-th row, the final reading is 
\begin{align}
    \alpha + \# \left\{ i \mid n_\alpha - \lambda^{(\alpha)}_i < j, \ i = 1,\dots, N-\alpha \right\}.
\end{align}
The $j$-th square in the $(\alpha+1)$-th row will have the reading
\begin{align}
    \alpha + 1 + \# \left\{ i \mid n_{\alpha+1} - \lambda^{(\alpha+1)}_i < j, \ i = 1,\dots, N-\alpha \right\}.
\end{align}
The constraint on the instanton configuration \eqref{eq:Young-condi} ensures that 
\begin{align}
    \# \left\{ i \mid n_\alpha - \lambda^{(\alpha)}_i < j, \ i = 1,\dots, N-\alpha \right\} \leq \# \left\{ i \mid n_{\alpha+1} - \lambda^{(\alpha+1)}_i < j, \ i = 1,\dots, N-\alpha \right\}
\end{align}
The reading of the squares in Young Tableaux $\bT_\bn[\boldsymbol\lambda]$ is always non-decreasing when moving rightward and always strictly increasing when moving downward. Thus $\bT_\bn[\boldsymbol\lambda]$ is semi-standard for any instanton configuration $\boldsymbol\lambda$. 

The ensemble in \eqref{eq:inst-quant} now sums over the Young Tableaux $\bT_\bn=(n_1,\dots,n_N)$ of shape $\bn$. We denote the reading of the $j$-th square in the $\alpha$-th row as $T_{\alpha j}$. These reading can be translated to the corresponding instanton configuration by
\begin{align}
    \lambda^{T,(\alpha)}_{n_\alpha+1-j} = T_{\alpha,j} - \alpha 
\end{align}
The $\lambda^{T,(\alpha)}$ denotes the conjugation of the instanton configuration $\lambda^{(\alpha)}$ (also known as the transpose of $\lambda^{(\alpha)}$). By the construction of Young Tableaux $\bT_\bn$, we have
\begin{align}
    T_{\alpha, j} \leq N.
\end{align}

The counting of instanton configuration $\boldsymbol\lambda$, which is the power of $z_\alpha$, is 
\begin{align}
    n_\alpha + \sum_{\beta<\alpha} \lambda_{\alpha-\beta}^{(\beta)} - \sum_{\beta<\alpha+1} \lambda_{\alpha+1-\beta}^{(\beta)} = n_\alpha - \lambda^{(\alpha)}_1 + \sum_{\beta<\alpha} \lambda^{(\beta)}_{\alpha-\beta} - \lambda_{\alpha+1-\beta}^{(\beta)} =: t_\alpha. 
\end{align}
We now argue that this value is the weight $t_\alpha$ of $\bT_\bn[\boldsymbol\lambda]$. In other words, $t_\alpha$ counts the occurrences of the number of $\alpha$ in $\bT_\bn[\boldsymbol\lambda]$. The first two terms are straight forward, it is the number of the number $\alpha$ after we increase the reading of squares in the $\alpha$-th column. 
In the $\beta$-th column with $\beta<\alpha$, a square that can have a reading of $\alpha$ needs to be increased $\alpha-\beta$ times but not any further. Since each $\lambda^{(\alpha)}_i$ only increase the reading by 1, The $\beta$-th column will have exactly $\lambda^{(\beta)}_{\alpha-\beta} - \lambda^{(\alpha)}_{\alpha+1-\beta}$ readings of the number $\alpha$.

The wave function \eqref{eq:inst-quant} can be rewritten as an ensemble over the semi-standard Young tableaux $\bT_\bn$ whose reading at the $(\alpha,j)$ square satisfy 
\begin{align}\label{eq:YT-condi}
    \alpha \leq T_{\alpha ,j} \leq T_{\alpha,j+1} \leq N, \ T_{\alpha,j} < T_{\alpha+1,j}.
\end{align}

Given a Young Tableaux $\bT_\bn[\boldsymbol\lambda]$ whose largest reading is less or equal to $N$ (not necessary equal to $N$).  We can define a series of sub Young Tableaux 
\begin{align}
    \varnothing = \bT_\bn^{(0)} [\boldsymbol\lambda] \subset \bT_\bn^{(1)} [\boldsymbol\lambda] \subset \cdots \subset \bT_\bn^{(N)} [\boldsymbol\lambda] = \bT_\bn [\boldsymbol\lambda]
\end{align}
The sub Young Tableaux $\bT_\bn^{(i)} [\boldsymbol\lambda] = \bn^{(i)} = (n_1^{(i)},n_2^{(i)},\dots,n_N^{(i)})$ has its reading less or equal to $i$. By its construction 
\begin{align}
    n^{(i)}_j = 0
\end{align}
if $j>i$.
The instanton configuration $\boldsymbol\lambda$ can be obtained by
\begin{align}
    \lambda^{(\alpha)}_k = n_\alpha - n^{(\alpha+k-1)}_\alpha.
\end{align}
The weight $t_\alpha$ of the Young Tableaux $\bT_{\bn}[\boldsymbol\lambda]$ equals 
\begin{align}
    t_\alpha = |\bn^{(\alpha)}|-|\bn^{(\alpha-1)}|.
\end{align}
where 
\begin{align}
    |\bn^{(\alpha)}| = \sum_{j=1}^{\alpha} n_{j}^{(\alpha)}.
\end{align}
The wave function \eqref{eq:instdef} from the defect instanton partition function can be now written in terms of ensemble over the Young Tableaux
\begin{align}\label{eq:Psi-comb}
    \Psi = \sum_{\bT_\bn} z^{\bT_\bn} \psi_{\bT_\bn}
\end{align}
where $z^{\bT_\bn} = \prod_{j=1}^N z_i^{t_i}$ and
\begin{align}
\begin{split}
    \psi_{\bT_\bn} & = \prod_{\o=2}^N \prod_{1\leq\beta<\alpha<\omega} \prod_{j=1}^{n^{(\omega)}_{\beta} - n_{\beta}^{(\omega-1)}} 
    \frac{\kappa(\beta-\alpha-1)+n_\alpha^{(\o-1)} - n_\beta^{(\o)} + j }{ \kappa(\beta-\alpha) + n_{\alpha}^{(\o-1)} - n_{\beta}^{(\o)} + j-1 } \times \frac{\kappa(\beta-\alpha)+n_{\alpha+1}^{(\o)} - n_\beta^{(\o)} + j-1 }{ \kappa(\beta-\alpha-1) + n_{\alpha+1}^{(\o)} - n_{\beta}^{(\o)} + j }
    \nonumber \\
    & = \prod_{\o=2}^N \prod_{1\leq\beta<\alpha<\omega} \frac{\Gamma(\kappa(\beta-\alpha-1)+n_\alpha^{(\o-1)}-n_\beta^{(\o-1)}+1)}{\Gamma (\kappa(\beta-\alpha-1) + n^{(\o-1)}_{\alpha} - n_\beta^{(\o)})} \frac{\Gamma(\kappa(\beta-\alpha)+n_\alpha^{(\o-1)}-n_\beta^{(\o)} -1 )}{ \Gamma (\kappa(\beta-\alpha) + n_\alpha^{(\o-1)} - n_\beta^{(\o-1)} ) } \\
    & \qquad \qquad \qquad \times \frac{\Gamma (\kappa(\beta-\alpha) + n_{\alpha+1}^{(\o)} - n_\beta^{(\o-1)} ) }{ \Gamma (\kappa (\beta-\alpha) + n_{\alpha+1}^{(\o)} - n_\beta^{(\o)} - 1 ) } \frac{\Gamma (\kappa(\beta-\alpha-1) + n_{\alpha+1}^{(\o)} - n_\beta^{(\o)}) }{\Gamma (\kappa(\beta-\alpha-1) + n_{\alpha+1}^{(\o)} - n_{\beta}^{(\o-1)} + 1 )}
    \, .
\end{split}
\end{align}
Eq.~\eqref{eq:Psi-comb} is the combinatorial formula for Jack polynomial~\cite[Chapter VI, \S10]{macdonald1998symmetric}.



In the massless limit $m \to 0$, which translates to $\kappa \to 1$ limit (Schur limit) of the Jack parameter under the Bethe/gauge correspondence, all instanton configuration $\boldsymbol\lambda$ satisfying \eqref{eq:Young-condi} shares a common pseudo-measure in the ensemble
\begin{align}
    \lim_{\nu \to 0} \CalZ_{\rm defect}[\boldsymbol\lambda] = 1.
\end{align}
The wave function $\Psi$ is an ensemble over the instanton configuration that satisfies \eqref{eq:Young-condi}. We immediately identify $\Psi$ as Schur polynomial using the Young Tableaux representation: 
\begin{align}
    \lim_{\kappa \to 0} \Psi 
    = \sum_{\bT_\bn} z_1^{t_1} \cdots z_{N}^{t_N} = s_\bn (z_1,\dots,z_{N}).
\end{align}

\subsubsection*{Example:} Let $N=3$, $\bn ={\tiny \Yvcentermath1 \yng(2,1)} =(2,1,0)$. We start with a Young Tableaux $\bT_\bn[\varnothing]$ that represents the no instanton configuration $\boldsymbol\lambda=(\varnothing,\varnothing,\varnothing)$
\begin{align}
    \bT_\bn [\varnothing] = \begin{ytableau}
    1 & 1 \cr 2
    \end{ytableau}
\end{align}
Here we will list out all Young Tableaux denoting the instanton configuration $\boldsymbol\lambda$. 
\begin{align}
    \bT_\bn [({\tiny \vc \yng(1), \varnothing,\varnothing})] & = \begin{ytableau}
    1 & 2 \cr 2
    \end{ytableau},  & 
    \bT_\bn [({\tiny \vc \yng(1,1), \varnothing,\varnothing})] & = \begin{ytableau}
    1 & 3 \cr 2
    \end{ytableau}, &
    \bT_\bn [({\tiny \vc \varnothing, \yng(1), \varnothing })] & = \begin{ytableau}
    1 & 1 \cr 3
    \end{ytableau}, \nonumber\\
    \bT_\bn [({\tiny \vc \yng(1), \yng(1) ,\varnothing})] & = \begin{ytableau}
    1 & 2 \cr 3
    \end{ytableau}, & 
    \bT_\bn [({\tiny \vc \yng(1,1), \yng(1) ,\varnothing})] & = \begin{ytableau}
    1 & 3 \cr 3
    \end{ytableau}, && \nonumber\\
    \bT_\bn [({\tiny \vc \yng(2), \yng(1) ,\varnothing})] & = \begin{ytableau}
    2 & 2 \cr 3
    \end{ytableau}, & \bT_\bn [({\tiny \vc \yng(2,1), \yng(1) ,\varnothing})] & = \begin{ytableau}
    2 & 3 \cr 3
    \end{ytableau}. &&
\end{align}
There are only eight semi-standard Young Tableaux of shape $\bn$, each of them corresponds to an instanton configuration. One can check that in each case the weight of $\bT_\bn[\boldsymbol\lambda]$ equals to the instanton counting.  


\subsection{Higher dimensions}

We have discussed the surface defect in four dimensional gauge theory, and its relation to the Jack polynomial.
From the gauge theory point of view, one can generalize this setup to higher dimensions.
Imposing codimension two defects, we would obtain 5d/3d and 6d/4d coupled systems, which correspond to the hierarchy of rational/trigonometric/elliptic integrable systems.
Based on a similar setup in five dimensions, it has been shown that the defect partition function can be identified with the Macdonald polynomial, which is an eigenfunction of the Ruijsenaars--Schneider operator~\cite{Koroteev:2018isw,Koroteev:2019byp}.
In fact, the Macdonald polynomial also has the tableau formula, which is a trigonometric analog of \eqref{eq:Psi-comb} (See \cite[Chapter VI, \S7]{macdonald1998symmetric}%
\footnote{%
See also \url{https://www.symmetricfunctions.com/macdonaldP.htm} 
} and \cite{Noumi2012direct}),
\begin{align}
    \psi_{\bT_\bn} & = \prod_{\o=2}^N \prod_{1\leq\beta<\alpha<\omega} \prod_{j=1}^{n^{(\omega)}_{\beta} - n_{\beta}^{(\omega-1)}} 
    \frac{1 - q^{n_\alpha^{(\o-1)} - n_\beta^{(\o)} + j} t^{\beta-\alpha-1} }{ 1 - q^{n_{\alpha}^{(\o-1)} - n_{\beta}^{(\o)} + j-1} t^{\beta-\alpha} }  \frac{1 - q^{n_{\alpha+1}^{(\o)} - n_\beta^{(\o)} + j-1} t^{\beta-\alpha} }{ 1 - q^{n_{\alpha+1}^{(\o)} - n_{\beta}^{(\o)} + j} t^{\beta-\alpha-1} }
    \nonumber \\ &
    = \prod_{\o=2}^N \prod_{1\leq\beta<\alpha<\omega} \frac{(q^{n_\alpha^{(\o-1)} - n_\beta^{(\o)} + 1} t^{\beta-\alpha-1};q)_\infty}{(q^{n_\alpha^{(\o-1)} - n_\beta^{(\omega-1)} + 1} t^{\beta-\alpha-1};q)_\infty}
    \frac{(q^{n_\alpha^{(\o-1)} - n_\beta^{(\omega-1)}} t^{\beta-\alpha};q)_\infty}{(q^{n_\alpha^{(\o-1)} - n_\beta^{(\o)}} t^{\beta-\alpha};q)_\infty}
    \nonumber \\ &
    \qquad \qquad \qquad \times 
    \frac{(q^{n_{\alpha+1}^{(\o)} - n_\beta^{(\o)}} t^{\beta-\alpha};q)_\infty}{(q^{n_{\alpha+1}^{(\o)} - n_\beta^{(\omega-1)}} t^{\beta-\alpha};q)_\infty}
    \frac{(q^{n_{\alpha+1}^{(\o)} - n_\beta^{(\omega)}+1} t^{\beta-\alpha-1};q)_\infty}{(q^{n_{\alpha+1}^{(\o)} - n_\beta^{(\o-1)}+1} t^{\beta-\alpha-1};q)_\infty}
    \, .
    \label{eq:tMacdonald}
\end{align}
The previous formula \eqref{eq:Psi-comb} is reproduced by putting $t = q^\kappa$ and then taking the limit $q \to 1$.
This expression is obtained in parallel from the defect partition function by replacing the index functor \eqref{eq:index_func} with the trigonometric version (5d/3d theory convention) although its derivation from the $qq$-character would be more involved.
We obtain an elliptic analog of the formula~\eqref{eq:tMacdonald} from the 6d/4d setup with the elliptic index,
\begin{align}
    \psi_{\bT_\bn} & = \prod_{\o=2}^N \prod_{1\leq\beta<\alpha<\omega} \prod_{j=1}^{n^{(\omega)}_{\beta} - n_{\beta}^{(\omega-1)}} 
    \frac{\theta(q^{n_\alpha^{(\o-1)} - n_\beta^{(\o)} + j} t^{\beta-\alpha-1};p) }{\theta( q^{n_{\alpha}^{(\o-1)} - n_{\beta}^{(\o)} + j-1} t^{\beta-\alpha};p) }  \frac{\theta( q^{n_{\alpha+1}^{(\o)} - n_\beta^{(\o)} + j-1} t^{\beta-\alpha};p) }{ \theta( q^{n_{\alpha+1}^{(\o)} - n_{\beta}^{(\o)} + j} t^{\beta-\alpha-1};p) }
    \nonumber \\
    & = \prod_{\o=2}^N \prod_{1\leq\beta<\alpha<\omega}
    \frac{\Gamma(q^{n_\alpha^{(\o-1)} - n_\beta^{(\omega-1)} + 1} t^{\beta-\alpha-1};q,p)}{\Gamma(q^{n_\alpha^{(\o-1)} - n_\beta^{(\o)} + 1} t^{\beta-\alpha-1};q,p)}
    \frac{\Gamma(q^{n_\alpha^{(\o-1)} - n_\beta^{(\o)}} t^{\beta-\alpha};q,p)}{\Gamma(q^{n_\alpha^{(\o-1)} - n_\beta^{(\omega-1)}} t^{\beta-\alpha};q,p)}
    \nonumber \\ &
    \qquad \qquad \qquad \times 
    \frac{\Gamma(q^{n_{\alpha+1}^{(\o)} - n_\beta^{(\omega-1)}} t^{\beta-\alpha};q,p)}{\Gamma(q^{n_{\alpha+1}^{(\o)} - n_\beta^{(\o)}} t^{\beta-\alpha};q,p)}
    \frac{\Gamma(q^{n_{\alpha+1}^{(\o)} - n_\beta^{(\o-1)}+1} t^{\beta-\alpha-1};q,p)}{\Gamma(q^{n_{\alpha+1}^{(\o)} - n_\beta^{(\omega)}+1} t^{\beta-\alpha-1};q,p)}
    \, .
    \label{eq:eMacdonald}
\end{align}
where $\Gamma(z;q,p)$ is the elliptic $\Gamma$-function~\eqref{eq:ell_gamma}.
This defines an elliptic analog of the Macdonald polynomial~\cite{Fukuda:2020czf,Awata:2020xfq}.

\section{Quantum Hall States} \label{sec:gauge to Hall}


In this section, we discuss a possible connection between four dimensional gauge theory and two dimensional fractional quantum Hall (FQH) effect.
The idea is as follows.
On the four dimensional gauge theory side, we apply the $\Omega$-background for each complex plane, which plays a role of the background $U(1)$ magnetic field.
See, e.g., \cite{Nekrasov:2002qd,Hama:2012bg}.
Then, imposing the surface defect and taking the bulk decoupling limit, we may focus on the two dimensional system with the background field, which realizes the QH effect.
In fact, it has been shown by Bernevig--Haldane~\cite{Bernevig:2007nek,Bernevig:2008rda,Bernevig:2009zz} that the wide class of FQH wave functions for the ground state and with the quasi-hole excitations are realized using the Jack polynomial (multiplying a trivial Gaussian factor which we will drop here) that we have already obtained from the gauge theory analysis. 
We pursue the direction suggested by Nekrasov~\cite{Nekrasov:2019AMJ} for more details, and construct more generic FQH wave functions.%
\footnote{%
We also remark that a gauge theory realization of the Moore-Read state has been discussed based on the AGT relation~\cite{Santachiara:2010bt}.
}

\subsection{Laughlin State}

The lowest Landau level (LLL) wave function is in general given by a product of conformally-invariant holomorhic multi-variable polynomial $\psi_\text{LLL}(z_1,\dots,z_N)$ and a trivial Gaussian factor that we will drop here. 
The Laughlin wave function is a key for understanding the physics of the FQH effect. It models the simplest abelian FQH state and is the building block to more general cases, both abelian and non-abelian ones. 
The Laughlin wave function of filling fraction $\frac{1}{r}$
\footnote{%
Here we consider the bosonic FQH states, which are directly related to the Jack polynomial.
Hence the parameter $r$ is even.
One can obtain the fermionic FQH states by multiplying the free fermion factor, $\prod_{\alpha < \beta}(z_\alpha - z_\beta)$. 
See~\cite{Bernevig:2007nek,Bernevig:2008rda,Bernevig:2009zz} for details. 
}
\begin{align}\label{def:Laughlin}
    \psi_\text{L}^{(r)} = \prod_{\alpha<\beta}(z_\alpha - z_\beta ) ^r = J ^{\frac1\kappa } _ {\bn^0(1,r)}, \ 
\end{align}
is the eigenfunction of Laplace-Beltrami operator \eqref{def:LBop} with the parameter identification
\begin{align}
    \kappa = - \frac{r-1}{2}. 
\end{align}
This can be easily verified by noticing the Laughlin state $\psi_\text{L}^{(r)}$ is annihilated by the Dunkl operator
\begin{align}
    D_\alpha^{(r)} = \frac{\p}{\p z_\alpha} - r \sum_{\beta\neq \alpha} \frac{1}{z_\alpha-z_\beta}.
\end{align}
Hence the Laughlin state is annihilated by $\sum_i z_i D_i^{(1)} z_i D_i^{(r)}$, which we identify
\begin{align}
    \sum_\alpha z_\alpha D_\alpha^{(1)} z_\alpha D_\alpha^{(r)} = \EH_{\text{LB}} - \frac{r}{12} N(N-1)(N+1+3r(N-1))
\end{align}
i.e. 
\begin{align}
    \EH_{\text{LB}} \psi_\text{L}^{(r)} = \frac{r}{12} N(N-1)(N+1+3r(N-1)) \psi_\text{L}^{(r)} . 
\end{align}

For Laughlin state to be a polynomial in $(z_\alpha)_{\alpha=1,\ldots,N}$, i.e. $r\in \BZ_{>0}$, it would require the Laplace-Beltrami coupling $\kappa$ to be negative. On the gauge theory side, that means the adjoint mass must be negative half integer
\begin{align}
    \kappa=-\frac{r-1}{2}.
\end{align}
We impose quantization condition \eqref{eq:quant} 
\begin{align}
    - \frac{a_\omega}{\ve_1} + \kappa \rho_\omega = (N-\omega) r , \ \omega = 1,\dots,N
\end{align}
such that the Young diagram $\bn$ is $(1,r,N)$-admissible. Indeed, we find the eigenvalue of gauge theory instanton partition function becomes
\begin{align}
    \EH_{\rm LB} \Psi = \frac{r}{12} N(N-1)(N+1+3r(N-1)) \Psi.
\end{align}
matching with that of Laughlin state. 

\subsection{Moore--Read state}

Let us turn our attention to the Moore-Read (MR) state~\cite{Moore:1991ks}. It was introduced as a model to study the FQH state with the filling fraction $\nu = \frac{5}{2}$, which is the $M=1$ case of 
\begin{align}\label{def:Moore-Read}
    \Psi_\text{MR}^{(M)}(z_1, \ldots, z_N) = \operatorname{Pf}\left( \frac{1}{z_\alpha - z_\beta} \right) \prod_{\alpha<\beta}^N (z_\alpha - z_\beta)^{M+1}.
\end{align}
Hereafter we consider the odd $M$ case, which is the bosonic analog of the MR state.
The Pfaffian factor in the Moore-Read state is annihilated by \cite{belavin1984infinite}
\begin{align}
    D^\text{Pf}_\alpha = \frac{\p^2}{\p z_\alpha^2} + \sum_{\beta \neq \alpha} \frac{4/3}{z_\alpha-z_\beta} \frac{\p}{\p z_\beta} + \frac{2/3}{(z_\alpha-z_\beta)^2}, \ D^\text{Pf}_\alpha \operatorname{Pf} \left( \frac{1}{z_\alpha - z_\beta} \right) = 0.
\end{align}
It's not hard to find that the first Moore--Read state, obtained by multiplying the Pfaffian with Vandermonde determinant, obeys
\begin{align}\label{eq:MR-spectrum}
    \EH_\text{LB}(\kappa^{-1}=-3) \Psi_\text{MR}^{(0)} = \frac{N(N+2)(5N+8)}{18} \Psi_\text{MR}^{(0)}. 
\end{align}

\paragraph{}
We consider the following quantization condition in the gauge theory:
\begin{align}
    n_{2\alpha-1} = n_{2\alpha} = N - 2\alpha, \ \alpha = 1,2,\dots,\frac{N}{2}.
\end{align}
The partition $\bn$ is $(2,2)$-admissible. The defect partition function $\Psi$ is an eigenfunction of the Laplace-Beltrami operator \eqref{eq:Schrodinger-2} with eigenvalue
\begin{align}
\begin{split}
    E(\bn) 
    & = \sum_{\alpha=1}^{\frac{N}{2}} \left( N-2\alpha + \kappa \left( \frac{N+1}{2} - 2\alpha \right)  \right)^2 + \left( N-2\alpha + \kappa \left( \frac{N+1}{2} - 2\alpha + 1 \right)  \right) - \sum_{\alpha'=1}^{N} \kappa^2 \left( \frac{N+1}{2} - \alpha' \right) \\
    & = \frac{N}{6} \left( 2(N+1)(N+2) + \kappa (N^2-4) \right)
\end{split}
\end{align}
Choosing $\kappa = -\frac{1}{3}$ reproduces the correct spectrum for the Moore-Read state in \eqref{eq:MR-spectrum}:
\begin{align}
    E(\bn) = \frac{N(N+2)(5N+8)}{18}.
\end{align}

\paragraph{Example:}
The $M=0$, $N=4$ Moore-Read state by \eqref{def:Moore-Read} is
\begin{align}
\begin{split}
    \Psi_\text{MR}^{(1)}(z_1,z_2,z_3,z_4) = 
    & \ z_1^2 z_2^2 + z_1^2 z_3^2 + z_1^2 z_4^2 + z_2^2 z_3^2 + z_2^2 z_4^2 + z_3^2 z_4^2 \\
    & - z_1^2 (z_2 z_3 + z_2 z_4 + z_3 z_4) - z_2^2 (z_1 z_3 + z_1 z_4 + z_3 z_4) \\
    & - z_3^2 (z_1 z_2 + z_1 z_4 + z_2 z_4) - z_4^2 (z_1 z_2 + z_1 z_3 + z_2 z_3) \\
    & + 6 z_1 z_2 z_3 z_4.
\end{split}
\end{align}

We consider the following quantization condition for the $U(4)$ gauge theory:
\begin{align}
    n_1 = n_2 = 2, \ n_3 = n_4 = 0.
\end{align}
The partition $\bn = {\tiny \vc \yng(2,2)}$ is $(2,2)$-admissible:
\begin{align}
    n_{i} - n_{i+2} \geq 2, \ i=0,1.
\end{align}
The instanton configuration that has non-zero pseudo measure must obey
\begin{align}
    \lambda_1^{(1)} \leq \lambda_1^{(3)} \leq 2, \ \lambda_2^{(2)} \leq \lambda_2^{(4)} \leq 2, \ \lambda_{3}^{(2)} = \lambda_{1}^{(4)} = 0
\end{align}
The $\lambda^{(2)}$ instanton configurations are of one of the following:
\begin{align}
    \varnothing, \ {\yng(1)}, \ \yng(2), \ \yng(1,1), \ \yng(2,1), \ \yng(2,2).
\end{align}
The $\lambda^{(1)}$ is always dominated by $\lambda^{(2)}$. Here we list out all the instanton configurations and their contribution toward the ensemble:
\renewcommand{\arraystretch}{1.3}
\begin{center}
    \begin{tabular}{|c||c|c|c|c|c|c|}
    \hline
    $\lambda^{(1)}$ ${\backslash}$ $\lambda^{(2)}$ & $\varnothing$ & {\tiny\Yvcentermath1\yng(1)} & {\tiny\Yvcentermath1\yng(2)} & {\tiny\Yvcentermath1\yng(1,1)} & {\tiny\Yvcentermath1\yng(2,1)} & {\tiny\Yvcentermath1\yng(2,2)} \\
    \hline\hline
    $\varnothing$ & $z_1^2z_2^2$ & $\frac{2\kappa}{\kappa+1} z_1^2z_2z_3$ & $z_1^2z_3^2$ & $\frac{2\kappa}{\kappa+1} z_1^2z_2z_4$ & $ \frac{2\kappa}{\kappa+1}z_1^2z_3z_4$ & $z_1^2z_4^2$ \\
    \hline
    {\tiny\Yvcentermath1\yng(1)} & & $-\frac{2\kappa}{\kappa+1} z_1z_2^2z_3$ & $\frac{2\kappa}{\kappa+1} z_1z_2z_3^2$ & $\frac{2\kappa}{\kappa+1} z_1z_2^2z_4$ & $ \frac{4\kappa^2}{(\kappa+1)^2} z_1z_2z_3z_4$ & $\frac{2\kappa}{\kappa+1} z_1z_2z_4^2$ \\
    \hline 
    {\tiny\Yvcentermath1\yng(2)} & & & $z_2^2z_3^2$ & & $\frac{2\kappa}{\kappa+1}z_2^2z_3z_4$ & $z_2^2z_4^2$ \\
    \hline
    {\tiny\Yvcentermath1\yng(1,1)} & & & & $ \frac{4\kappa^2}{(\kappa+1)^2} \frac{\kappa+2}{2\kappa+1} z_1z_2z_3z_4$ & $\frac{2\kappa}{\kappa+1} z_1 z_3^2 z_4$ & $\frac{2\kappa}{\kappa+1}z_1z_3z_4^2$ \\
    \hline 
    {\tiny\Yvcentermath1\yng(2,1)} & & & & & $\frac{2\kappa}{\kappa+1}z_2z_3^2z_4$ & $\frac{2\kappa}{\kappa+1}z_2z_3z_4^2$\\
    \hline 
    {\tiny\Yvcentermath1\yng(2,2)} & & & & & & $z_3^2z_4^2$\\
    \hline
\end{tabular}
\end{center}
\renewcommand{\arraystretch}{1}

As we have demonstrated earlier, each instanton configuration $\boldsymbol\lambda = (\lambda^{(1)},\lambda^{(2)},\lambda^{(3)}=\varnothing)$ can be expressed using semi-standard Young Tableaux.  We start with
\begin{align}
    \bT_\bn [\varnothing] = \begin{ytableau}
    1 & 1 \cr 2 & 2
    \end{ytableau}
\end{align}
The 20 instanton configurations can be captured by all semi-standard Young Tableaux $\bT_\bn$:
\begin{align}
\begin{split}
    & \bT_\bn \left[ \left( {\tiny \varnothing, \varnothing} \right) \right]= \begin{ytableau}
    1 & 1 \cr 2 & 2
    \end{ytableau}, \  
    \bT_\bn \left[ \left( {\tiny \vc \varnothing, \yng(1)}\right) \right] = \begin{ytableau}
    1 & 1 \cr 2 & 3
    \end{ytableau}, \ 
    \bT_\bn \left[ \left( {\tiny \vc \varnothing, \yng(2)} \right) \right] = \begin{ytableau}
    1 & 1 \cr 3 & 3
    \end{ytableau}, \ 
    \bT_\bn \left[ \left( {\tiny \vc \yng(1), \yng(1)} \right) \right] = \begin{ytableau}
    1 & 2 \cr 2 & 3
    \end{ytableau}, \\
    & \bT_\bn \left[ \left( {\tiny \vc \yng(1), \yng(2)} \right) \right] = \begin{ytableau}
    1 & 2 \cr 3 & 3
    \end{ytableau}, \ 
    \bT_\bn \left[ \left( {\tiny \vc \yng(2), \yng(2)} \right) \right] = \begin{ytableau}
    2 & 2 \cr 3 & 3
    \end{ytableau}, \ 
    \bT_\bn \left[ \left( {\tiny \vc \varnothing, \yng(1,1)} \right) \right] = \begin{ytableau}
    1 & 1 \cr 2 & 4
    \end{ytableau}, \ 
    \bT_\bn \left[ \left( {\tiny \vc \yng(1), \yng(1,1)} \right) \right]=  \begin{ytableau}
    1 & 2 \cr 2 & 4
    \end{ytableau}, \\
    & \bT_\bn \left[ \left( {\tiny \vc \yng(1,1), \yng(1,1)} \right) \right] = \begin{ytableau}
    1 & 3 \cr 2 & 4
    \end{ytableau}, \ 
    \bT_\bn \left[ \left( {\tiny \vc \varnothing, \yng(2,1)} \right) \right]= \begin{ytableau}
    1 & 1 \cr 3 & 4
    \end{ytableau}, \ 
    \bT_\bn \left[ \left( {\tiny \vc \yng(1), \yng(2,1)} \right) \right]= \begin{ytableau}
    1 & 2 \cr 3 & 4
    \end{ytableau}, \ 
    \bT_\bn \left[ \left( {\tiny \vc \yng(1,1), \yng(2,1)} \right) \right]= \begin{ytableau}
    1 & 3 \cr 3 & 4
    \end{ytableau}, \\
    & \bT_\bn \left[ \left( {\tiny \vc \yng(2), \yng(2,1)} \right) \right]= \begin{ytableau}
    2 & 2 \cr 3 & 4
    \end{ytableau}, \ 
    \bT_\bn \left[ \left( {\tiny \vc \yng(2,1), \yng(2,1)} \right) \right]= 
    \begin{ytableau}
    2 & 3 \cr 3 & 4
    \end{ytableau}, \ 
    \bT_\bn \left[ \left( {\tiny \vc \varnothing, \yng(2,2)} \right) \right]=   \begin{ytableau}
    1 & 1 \cr 4 & 4
    \end{ytableau}, \
    \bT_\bn \left[ \left( {\tiny \vc \yng(1), \yng(2,2)} \right) \right]= \begin{ytableau}
    1 & 2 \cr 4 & 4
    \end{ytableau}, \\
    & \bT_\bn \left[ \left( {\tiny \vc \yng(1,1), \yng(2,2)} \right) \right]= \begin{ytableau}
    1 & 3 \cr 4 & 4
    \end{ytableau}, \ 
    \bT_\bn \left[ \left( {\tiny \vc \yng(2), \yng(2,2)} \right) \right]= \begin{ytableau}
    2 & 2 \cr 4 & 4
    \end{ytableau}, \ 
    \bT_\bn \left[ \left( {\tiny \vc \yng(2,1), \yng(2,2)} \right) \right]= \begin{ytableau}
    2 & 3 \cr 4 & 4
    \end{ytableau}, \ 
    \bT_\bn \left[ \left( {\tiny \vc \yng(2,2), \yng(2,2)} \right) \right]= \begin{ytableau}
    3 & 3 \cr 4 & 4
    \end{ytableau}. \nonumber
\end{split}
\end{align}
The readings of each square in every Young Tableaux satisfy
\begin{align}
    \alpha \leq T_{\alpha,j} \leq T_{\alpha,j+1} \leq N, \ T_{\alpha, j} < T_{\alpha+1, j}.
\end{align}

The instanton partition function for general value of $\kappa$ is given by
\begin{align}
\begin{split}
    \Psi(z_1,z_2,z_3,z_4) = 
    & \ z_1^2 z_2^2 + z_1^2 z_3^2 + z_1^2 z_4^2 + z_2^2 z_3^2 + z_2^2 z_4^2 + z_3^2 z_4^2 \\
    & +\frac{2\kappa}{\kappa+1} z_1^2 (z_2 z_3 + z_2 z_4 + z_3 z_4) + \frac{2\kappa}{\kappa+1} z_2^2 (z_1 z_3 + z_1 z_4 + z_3 z_4) \\
    & + \frac{2\kappa}{\kappa+1} z_3^2 (z_1 z_2 + z_1 z_4 + z_2 z_4) + \frac{2\kappa}{\kappa+1} z_4^2 (z_1 z_2 + z_1 z_3 + z_2 z_3) \\
    & + \frac{12\kappa^2}{(\kappa+1)(2\kappa+1)} z_1 z_2 z_3 z_4.
\end{split}
\end{align}
As we can see $\Psi$ obeys the $(2,2)$-admissible condition. It does not have pole at 
\begin{align}
    \kappa = - \frac{2-1}{2+1} = - \frac{1}{3}
\end{align}
In the $\kappa \to -\frac{1}{3}$ limit, the instanton partition function matches with the 
\begin{align}
    \left. \Psi (z_1,z_2,z_3,z_4) \right|_{\bn = {\tiny\Yvcentermath1\yng(2,2)} } = \Psi_\text{MR}^{(0)}(z_1,z_2,z_3,z_4) = J^{-\frac{1}{3}}_{\bn}(z_1,z_2,z_3,z_4).
\end{align}

\subsection{Admissible Condition}
Jack polynomials $J^{\frac{1}{\kappa}}_{\bn}$ are defined on an integer partition $\bn$ and a parameter $\kappa$. 
Let $k$, $r$ be positive integers such that $r\geq 2$ and $k+1$, $r-1$ are coprime.  
The partition $\bn$ is called \emph{$(k,r)$-admissible} if it satisfies
\begin{align}
    n_i - n_{i+k} \geq r, \ 1\leq i \leq N-k.
\end{align}
In general, Jack polynomial can have poles at a negative rational value of $\kappa$. However if the partition $\bn$ is $(k,r)$-admissible then the Jack polynomial will not have a pole at~\cite{Feigin:2002IMRN},%
\footnote{%
It has been discussed that a different parameter specialization provides a simplification of the wave function~\cite{Sokolov:2014kma}.
}
\begin{align}
    \kappa = - \frac{r-1}{k+1}.
    \label{kappa_admissible}
\end{align}
This is known as the \emph{admissible condition}. The admissible condition considers the pole structure at a handful of particular values of Jack parameter $\kappa$. It does not mean the Jack polynomial is entire function for $\kappa$. 

It has been pointed out~\cite{Bernevig:2007nek,Bernevig:2008rda,Bernevig:2009zz} that this admissible condition properly captures the clustering property of the FQH state, so that the corresponding wave function is generally obtained as the Jack polynomial with the negative couplings.
For example, the Laughlin state and the MR state correspond to $k = 1$ and $2$, respectively.
The higher $k$ cases are the Read-Rezayi states, which are associated with $\mathbb{Z}_k$-parafermion CFT, while the $k = 2$ case corresponds to the Ising CFT.
One can also apply this formalism to the FQH states with spin degrees of freedom~\cite{Estienne:2011,Kimura:2012sc}.

We have seen in the example in the $N=2$ case with partition $\bn = {\tiny \yng(2)} = (2,0)$. This partition $\bn$ is $(1,2)$ admissible 
\begin{align}
    n_1-n_2 \geq 2. 
\end{align}
The Jack polynomial defined based on this partition
\begin{align}
    J_{(2)}^{\frac{1}{\kappa}}(z_1,z_2) = \frac{1}{1+\kappa} p_2 + \frac{\kappa}{1+\kappa} p_1^2
\end{align}
does not have pole at 
\begin{align}
    \kappa = -\frac{2-1}{1+1} = - \frac{1}{2}. 
\end{align}
Instead it has pole at $\kappa = -1$. This corresponds to the $(1,2)$-admissible condition, which $\bn$ does not obey. 

\subsubsection{Instanton sum formula}

In the previous section we express the Jack polynomial as ensemble over instanton configuration \eqref{eq:inst-quant}. 
It is easier to see the pole structure of the wave function $\Psi$ \eqref{def:eigen} by rewriting the pseudo measure in the form of $\Gamma$-functions by multiplying the 1-loop factor in \eqref{eq:1-loop}: 
\begin{align}
    \Psi = \prod_{\alpha=1}^N z_\alpha^{n_\alpha} \sum_{\hat\lambda} \prod_{\omega=2}^{N} \kq_{\omega-2} ^{\sum_{\alpha<\omega}\lambda_{\omega-\alpha}^{(\alpha)} } 
    & \prod_{\alpha,\beta<\omega} \frac{ \Gamma (n_\beta-n_\alpha + \kappa (\alpha-\beta) + \lambda_{\omega-\alpha}^{(\alpha)} - \lambda_{\omega-\beta}^{(\beta)} +1-\kappa ) }{\Gamma (n_\beta-n_\alpha + \kappa (\alpha-\beta) + \lambda_{\omega+1-\alpha}^{(\alpha)} - \lambda_{\omega-\beta}^{(\beta)} +1-\kappa )} \nonumber\\
    & \times \prod_{\alpha,\beta<\omega} \frac{\Gamma (n_\beta-n_\alpha + \kappa (\alpha-\beta) + \lambda_{\omega+1-\alpha}^{(\alpha)} - \lambda_{\omega-\beta}^{(\beta)})}{ \Gamma (n_\beta-n_\alpha + \kappa (\alpha-\beta) + \lambda_{\omega-\alpha}^{(\alpha)} - \lambda_{\omega-\beta}^{(\beta)}) } \nonumber\\
    & \times \prod_{\beta<\omega} \frac{\Gamma (n_\beta-n_\omega + \kappa (\omega-\beta) + \lambda_{1}^{(\omega)} - \lambda_{\omega-\beta}^{(\beta)})}{ \Gamma (n_\beta-n_\omega + \kappa (\omega-\beta) + \lambda_{1}^{(\omega)} - \lambda_{\omega-\beta}^{(\beta)} +1 - \kappa ) } \nonumber\\
    & \times \prod_{\beta<\omega} \frac{ \Gamma(n_\beta-n_\omega+\kappa(\omega-\beta)+1-\kappa) }{\Gamma( n_\beta-n_\omega + \kappa (\omega-\beta) )}.
\end{align}
The last line comes from the 1-loop factor and does not depend on instanton configuration $\boldsymbol\lambda$. In order for $\Psi$ to be finite, The pole coming form the 4 $\Gamma$-function must be canceled by $\Gamma$-functions in the denominators. 
Let $\kappa \to -\frac{s}{t}$ be a negative rational number. $s,t \in \BZ_{>0}$ are coprime. We isolated out the $\Gamma$-functions whose arguments are integers in the limit $\kappa \to -\frac{s}{t}$: 
\begin{align}\label{eq:admi-rel}
    & \prod_{\omega=2}^{N} \prod_{\beta+t+1<\omega} \frac{ \Gamma (n_\beta-n_{\beta+t+1} +\kappa t + \lambda_{\omega-\beta-t-1}^{(\beta+t+1)} - \lambda_{\omega-\beta}^{(\beta)} +1 ) }{\Gamma (n_\beta-n_{\beta+t+1} +\kappa t + \lambda_{\omega-\beta-t}^{(\beta+t+1)} - \lambda_{\omega-\beta}^{(\beta)} +1 ) } \nonumber\\
    & \times \prod_{\omega=2}^{N} \prod_{\beta+t-1<\omega} \frac{ \Gamma (n_\beta - n_{\beta+t-1} + \kappa t + \lambda_{\omega-\beta-t+1}^{(\beta+t-1)} - \lambda_{\omega-\beta+1}^{(\beta)} ) }{ \Gamma( n_\beta - n_{\beta+t-1} + \kappa t + \lambda_{\omega-\beta-t+1}^{(\beta+t-1)} - \lambda_{\omega-\beta}^{(\beta)} ) } \nonumber\\
    & \times \prod_{\omega=2}^{N} \prod_{\beta+t<\omega} \frac{\Gamma (n_\beta-n_{\beta+t} +\kappa t + \lambda_{\omega+1-\beta-t}^{(\beta+t)} - \lambda_{\omega-\beta}^{(\beta)}) }{ \Gamma (n_\beta-n_{\beta+t} +\kappa t + \lambda_{\omega-\beta-t}^{(\beta+t)} - \lambda_{\omega-\beta}^{(\beta)}) } \nonumber\\
    & \times \prod_{\omega=2}^{N} \prod_{\beta+t<\omega} \frac{ \Gamma (n_\beta-n_{\beta+t} +\kappa t + \lambda_{\omega-\beta-t}^{(\beta+t)} - \lambda_{\omega-\beta}^{(\beta)} + 1) }{\Gamma (n_\beta-n_{\beta+t} +\kappa t + \lambda_{\omega-\beta-t}^{(\beta+t)} - \lambda_{\omega-\beta+1}^{(\beta)} +1) } \nonumber\\
    & \times \prod_{\beta+t \leq N} \frac{\Gamma (n_\beta-n_{\beta+t} +\kappa t + \lambda_{1}^{(\beta+t)} - \lambda_{t}^{(\beta)})} {\Gamma( n_\beta-n_{\beta+t} +\kappa t )} \nonumber\\
    & \times \prod_{\beta+t+1\leq N} \frac{ \Gamma(n_\beta-n_{\beta+t+1} +\kappa t+1) } { \Gamma (n_\beta-n_{\beta+t+1}+\kappa t + \lambda_{1}^{(\beta+t+1)} - \lambda_{t+1}^{(\beta)} +1 ) }
\end{align}
The two terms that have no dependence on the instanton configuration $\boldsymbol\lambda$ comes from the 1-loop contribution $\CalZ_\text{1-loop}$. Their combined contribution is
\begin{align}
\begin{split}
    \prod_{\beta+t+1\leq N} \frac { \Gamma(n_\beta-n_{\beta+t+1}+\kappa t+1) } {\Gamma( n_\beta-n_{\beta+t} +\kappa t )} \times \frac{1}{\Gamma (n_{N-1-t} +\kappa t) }.
\end{split}
\end{align}
It's obvious it does not give poles as the $\Gamma$-function in the numerator has larger argument than the denominator counter part. 
For the $\Gamma$-functions in the first line, a similar argument holds
\begin{align}
    & n_\beta-n_{\beta+t+1} + \kappa t + \lambda_{\omega-\beta-t-1}^{(\beta+t+1)} - \lambda_{\omega-\beta}^{(\beta)} \geq n_\beta-n_{\beta+t+1} +\kappa t + \lambda_{\omega-\beta-t}^{(\beta+t+1)} - \lambda_{\omega-\beta}^{(\beta)} 
\end{align}

The numerator in the fourth line can be combined with denominator from the third line. It's obvious that it will not give any poles. 


At this stage the potential poles coming from the gamma function in the 1st, 4th, and 6th line in \eqref{eq:admi-rel} will be canceled by the zeros coming from the gamma function in the denominators in the 1st, 3rd, and 5th line in \eqref{eq:admi-rel}. 
This leave us with the numerator in the second and third line, and denominator in the second and fourth line
\begin{align}\label{eq:analysis}
\begin{split}
    & \prod_{\omega=2}^{N} \prod_{\beta+t \leq \omega} \frac{ \Gamma (n_\beta - n_{\beta+t-1} + \kappa t + \lambda_{\omega-\beta-t+1}^{(\beta+t-1)} - \lambda_{\omega-\beta+1}^{(\beta)} ) \Gamma (n_\beta-n_{\beta+t} +\kappa t + \lambda_{\omega+1-\beta-t}^{(\beta+t)} - \lambda_{\omega-\beta}^{(\beta)}) }{ \Gamma( n_\beta - n_{\beta+t-1} + \kappa t + \lambda_{\omega-\beta-t+1}^{(\beta+t-1)} - \lambda_{\omega-\beta}^{(\beta)} ) } \\
    & \times \prod_{\omega=2}^{N} \prod_{\beta+t<\omega} \frac{ (n_\beta-n_{\beta+t} +\kappa t + \lambda_{\omega-\beta-t}^{(\beta+t)} - \lambda_{\omega-\beta}^{(\beta)}) }{\Gamma (n_\beta-n_{\beta+t} +\kappa t + \lambda_{\omega-\beta-t}^{(\beta+t)} - \lambda_{\omega-\beta+1}^{(\beta)}+1) }  
\end{split}
\end{align}
The last line comes from the ratio between numerator in the fourth line and denominator in the third line.
We notice that the arguments of the four $\Gamma$-functions satisfy the relation
\begin{align}
\begin{split}
    & n_\beta - n_{\beta+t-1} + \kappa t + \lambda_{\omega-\beta-t+1}^{(\beta+t-1)} - \lambda_{\omega-\beta}^{(\beta)} \\
    & \leq n_\beta - n_{\beta+t-1} + \kappa t + \lambda_{\omega-\beta-t+1}^{(\beta+t-1)} - \lambda_{\omega-\beta+1}^{(\beta)}, \ n_\beta-n_{\beta+t} + \kappa t + \lambda_{\omega+1-\beta-t}^{(\beta+t)} - \lambda_{\omega-\beta}^{(\beta)} \\
    & \leq n_\beta-n_{\beta+t} +\kappa t + \lambda_{\omega-\beta-t}^{(\beta+t)} - \lambda_{\omega-\beta+1}^{(\beta)}
\end{split}
\end{align}
This tells us that with a fixed $(\beta,\omega)$, the Gamma factors can only have at most first order degree pole in the $\kappa \to -\frac{s}{t}$ limit.

Let us consider the case $\o=N$, which gives
\begin{align}
\begin{split}
    & \prod_{\beta+t \leq N} \frac{ \Gamma (n_\beta - n_{\beta+t-1} - s + \lambda_{N-\beta-t+1}^{(\beta+t-1)} ) \Gamma (n_\beta-n_{\beta+t} -s - \lambda_{N-\beta}^{(\beta)}) }{ \Gamma( n_\beta - n_{\beta+t-1} + -s  + \lambda_{N-\beta-t+1}^{(\beta+t-1)} - \lambda_{N-\beta}^{(\beta)} ) } \\
    & \times \prod_{\beta+t<N} \frac{ (n_\beta-n_{\beta+t} - s + \lambda_{N-\beta-t}^{(\beta+t)} - \lambda^{(\beta)}_{N-\beta} ) }{\Gamma (n_\beta-n_{\beta+t} -s + \lambda_{N-\beta-t}^{(\beta+t)} +1) }
\end{split}
\end{align}
In order to maximize the number of potential poles, we look at the instanton configuration $\boldsymbol\lambda$ that minimize the argument of $\Gamma$-functions in the numerator, which are 
\begin{subequations}
\begin{align}
    & n_\beta - n_{\beta+t-1} - s + \lambda_{N-\beta-t+1}^{(\beta+t-1)} \geq n_{\beta} - n_{\beta+t-1} - s; \\
    & n_\beta-n_{\beta+t} -s - \lambda_{N-\beta}^{(\beta)} \geq n_{\beta+1} - n_{\beta+t} - s.
\end{align}
\end{subequations}
The $\Gamma$-functions will be finite if the partition $\bn$ satisfy the admissible condition
\begin{align}
    n_\beta - n_{\beta+t-1} \geq s + 1, \ \beta = 1,\dots, N-t+1.
\end{align}
at $\kappa= -\frac{s}{t}$. 

The condition \eqref{eq:Young-condi-3} provides a tool to analyze the argument of these $\Gamma$-functions in \eqref{eq:analysis} 
\begin{align}
    \lambda_{\omega-\alpha}^{(\alpha)} \leq n_{\alpha} - n_{N+1+\alpha-\omega}, \ \lambda_{\omega+1-\alpha}^{(\alpha)} \leq n_{\alpha} - n_{N+\alpha-\omega}
\end{align}
for all $\alpha<\omega$. 
The number of $\Gamma$-function that has poles in \eqref{eq:analysis} is limited based on its structure. Let us consider the following case: When $\lambda^{(\beta+t-1)}_{1} = \lambda_{1}^{(\beta+t)} = 0$ such that the $\Gamma$-function in the numerator in \eqref{eq:analysis} will have the smallest argument.  And in order to have poles with $\beta$ fixed, we would line to maximize the instanton configuration. It becomes
\begin{align}
    \begin{split}
    & \prod_{\beta} \prod_{\omega=\beta+t}^{N} \frac{ \Gamma (n_\beta - n_{\beta+t-1} + \kappa t - \lambda_{\omega-\beta+1}^{(\beta)} ) }{ \Gamma( n_\beta - n_{\beta+t-1} + \kappa t - \lambda_{\omega-\beta}^{(\beta)} ) } 
    \times \prod_{\o=\beta+t+1}^N \frac{\Gamma (n_{\beta}-n_{\beta+t} +\kappa t - \lambda_{\omega-\beta}^{(\beta)}+1) }{\Gamma (n_\beta-n_{\beta+t} +\kappa t - \lambda_{\omega-\beta+1}^{(\beta)} +1) } \\
    & = \prod_{\beta=1}^{N-t} \frac{\Gamma(n_{\beta} - n_{\beta+t-1} + \kappa t )}{\Gamma(n_{\beta}-n_{\beta+t-1} + \kappa t - \lambda_{t}^{(\beta)} )} \times
    \prod_{\beta=1}^{N-t-1} \frac{ \Gamma(n_{\beta} - n_{\beta+t} + \kappa t - \lambda_{t+1}^{(\beta)} + 1 ) }{\Gamma(n_{\beta}-n_{\beta+t} + \kappa t + 1) }
\end{split}
\end{align}
The terms with instanton configuration $\lambda^{(\beta)}$ is finite in $\kappa \to -\frac{s}{t}$ limit according to \eqref{eq:Young-condi} by
\begin{align}
    n_{\beta+1}-n_{\beta+t} + \kappa t - \lambda^{(\beta+1)}_{t} \leq n_{\beta} - n_{\beta+t} + \kappa - \lambda^{(\beta)}_t \leq n_{\beta} - n_{\beta+t} + \kappa - \lambda^{(\beta)}_{t+1}
\end{align}
The admissible condition guarantees it does not have pole in the $\kappa \to -\frac{s}{t}$ limit for $\beta=1,2,\dots,t$. 


\subsubsection*{Example 1: }
Let $N=6$ and $\bn=(4,4,2,2,0,0)$. The partition satisfies the admissible condition
\begin{align}\label{eq:admi-(2,2)}
    n_i - n_{i+2} \geq 2.
\end{align}
The instanton configuration must obey
\begin{subequations}\label{eq:inst for N=6,(2,2)}
\begin{align}
    & \lambda^{(5)} = \lambda^{(6)} = \varnothing ;\\
    & \lambda_{i}^{(3)} \leq \lambda_{i}^{(4)} \leq 2, \ i = 1,2 ; \quad \lambda^{(3)}_3 = 0; \\
    & \lambda_{i}^{(1)} \leq \lambda^{(2)}_i \leq \lambda_i^{(3)} + 2, \ i = 1,2,3,4 ; \quad \lambda^{(1)}_5 = 0;\\
    & \lambda^{(\alpha)}_{7-\alpha} = 0.
\end{align}
\end{subequations}
The potential poles in \eqref{eq:analysis} comes from the $\Gamma$-function in the numerator:
\begin{align}
\begin{split}
    & \prod_{\omega=4}^6 \prod_{\beta+3 \leq \omega} \Gamma \left( n_\beta - n_{\beta+2} - 1 + \lambda^{(\beta+2)}_{\omega-\beta-2} - \lambda_{\omega-\beta+1}^{(\beta)} \right) \Gamma \left( n_\beta - n_{\beta+3} - 1 + \lambda^{(\beta+3)}_{\omega-\beta-2} - \lambda^{(\beta)}_{\omega-\beta} \right) \\
    = & {  \Gamma (1+ \lambda_1^{(3)} - \lambda_4^{(1)} ) } \Gamma (1+ \lambda_2^{(3)} - \lambda_5^{(1)} ) { \Gamma (1+ \lambda_1^{(4)} - \lambda_4^{(2)} ) } \Gamma (1+ \lambda_3^{(3)} - \lambda_6^{(1)} ) \Gamma (1+ \lambda_2^{(4)} - \lambda_5^{(2)} ) \Gamma (1+ \lambda_1^{(5)} - \lambda_4^{(3)} ) \\
    & \times \Gamma(1+\lambda^{(4)}_1 - \lambda^{(1)}_3 ) {\Gamma (1+ \lambda_2^{(4)} - \lambda_4^{(1)} )} \Gamma (1+ \lambda_3^{(4)} - \lambda_5^{(1)} ) \Gamma(3+\lambda^{(5)}_1 - \lambda^{(2)}_{3}) \Gamma ( 3 + \lambda_2^{(5)} - \lambda_4^{(2)} ) \Gamma (1+\lambda^{(6)}_1-\lambda^{(3)}_3) \\
    = & { \Gamma (1+ \lambda_1^{(3)} - \lambda_4^{(1)} ) } \Gamma (1+ \lambda_2^{(3)} ) { \Gamma (1+ \lambda_1^{(4)} - \lambda_4^{(2)} ) } \Gamma (1+ \lambda_3^{(3)} ) \Gamma (1+ \lambda_2^{(4)} ) \Gamma (1) \\
    & \times { \Gamma(1+\lambda^{(4)}_1 - \lambda^{(1)}_3 )} { \Gamma (1+ \lambda_2^{(4)} - \lambda_4^{(1)} )} \Gamma (1 ) \Gamma(3 - \lambda^{(2)}_{3}) \Gamma ( 3 - \lambda_4^{(2)} ) \Gamma (1)
\end{split}
\end{align}
Not all but only four four $\Gamma$-functions above that can have zero or negative argument:
\begin{align}
    \Gamma (1+ \lambda_1^{(3)} - \lambda_4^{(1)} ) \Gamma (1+ \lambda_1^{(4)} - \lambda_4^{(2)} ) \Gamma(1+\lambda^{(4)}_1 - \lambda^{(1)}_3 ) \Gamma (1+ \lambda_2^{(4)} - \lambda_4^{(1)} ).
\end{align}
The $\Gamma$-function in the denominator in the first line in \eqref{eq:analysis} is
\begin{align}
\begin{split}
    & \prod_{\omega=4}^6 \prod_{\beta+3 \leq \omega} \Gamma \left( n_\beta - n_{\beta+2} - 1 + \lambda^{(\beta+2)}_{\omega-\beta-2} - \lambda_{\omega-\beta}^{(\beta)} \right) \\
    & = { \Gamma (1+ \lambda_1^{(3)} - \lambda_3^{(1)} )  \Gamma (1+ \lambda_2^{(3)} - \lambda_4^{(1)} )  \Gamma (1+ \lambda_1^{(4)} - \lambda_3^{(2)} ) } \\
    & \quad \times \Gamma (1+ \lambda_3^{(3)} - \lambda_5^{(1)} ) { \Gamma (1+ \lambda_2^{(4)} - \lambda_4^{(2)} ) } \Gamma (1+ \lambda_1^{(5)} - \lambda_3^{(3)} ) \\
    & = \Gamma (1+ \lambda_1^{(3)} - \lambda_3^{(1)} ) \Gamma (1+ \lambda_2^{(4)} - \lambda_4^{(2)} ) \Gamma (1+ \lambda_1^{(4)} - \lambda_3^{(2)} ) \Gamma (1+ \lambda_2^{(3)} - \lambda_4^{(1)} ) \Gamma (1) \Gamma (1)
\end{split}
\end{align}
All the potential poles coming from the $\Gamma$-function in the numerator of \eqref{eq:analysis} will be canceled by the $\Gamma$-function in the denominator as the instanton configuration must obey \eqref{eq:inst for N=6,(2,2)}. Here we see the Jack polynomial of $(2,2)$-admissible partition $\bn=(4,4,2,2,0,0)$ does not have a pole at $\kappa \to - \frac{1}{3}$.

\subsubsection*{Example 2: }
Next we consider a slightly bigger partition $\bn=(5,4,3,2,1,0)$ which satisfies the same admissible condition \eqref{eq:admi-(2,2)}. The instanton configuration must obey 
\begin{align}
    \lambda^{(\alpha+1)}_{\omega-\alpha} + 1 \geq \lambda_{\omega-\alpha}^{(\alpha)}, \quad \lambda^{(\alpha)}_{7-\alpha}=0.
\end{align}
The $\Gamma$-functions in the numerator of \eqref{eq:analysis} in the limit $\kappa \to -\frac{1}{3}$ are 
\begin{align}
\begin{split}
    & \prod_{\omega=4}^6 \prod_{\beta+3 \leq \omega} \Gamma \left( n_\beta - n_{\beta+2} - 1 + \lambda^{(\beta+2)}_{\omega-\beta-2} - \lambda_{\omega-\beta+1}^{(\beta)} \right) \Gamma \left( n_\beta - n_{\beta+3} - 1 + \lambda^{(\beta+3)}_{\omega-\beta-2} - \lambda^{(\beta)}_{\omega-\beta} \right) \\
    = & {  \Gamma (1+ \lambda_1^{(3)} - \lambda_4^{(1)} ) } \Gamma (1+ \lambda_2^{(3)} - \lambda_5^{(1)} ) { \Gamma (1+ \lambda_1^{(4)} - \lambda_4^{(2)} ) } \Gamma (1+ \lambda_3^{(3)} - \lambda_6^{(1)} ) \Gamma (1+ \lambda_2^{(4)} - \lambda_5^{(2)} ) \Gamma (1+ \lambda_1^{(5)} - \lambda_4^{(3)} ) \\
    & \times \Gamma(2+\lambda^{(4)}_1 - \lambda^{(1)}_3 ) {\Gamma (2+ \lambda_2^{(4)} - \lambda_4^{(1)} )} \Gamma (2+ \lambda_3^{(4)} - \lambda_5^{(1)} ) \Gamma(2+\lambda^{(5)}_1 - \lambda^{(2)}_{3}) \Gamma ( 2 + \lambda_2^{(5)} - \lambda_4^{(2)} ) \Gamma (2+\lambda^{(6)}_1-\lambda^{(3)}_3) 
\end{split}
\end{align}
The $\Gamma$-functions that can have zero or negative argument are
\begin{align}
\begin{split}
    & {  \Gamma (1+ \lambda_1^{(3)} - \lambda_4^{(1)} ) } \Gamma (1+ \lambda_2^{(3)} - \lambda_5^{(1)} ) { \Gamma (1+ \lambda_1^{(4)} - \lambda_4^{(2)} ) } \\
    & \times \Gamma(2+\lambda^{(4)}_1 - \lambda^{(1)}_3 ) {\Gamma (2+ \lambda_2^{(4)} - \lambda_4^{(1)} )} \Gamma(2+\lambda^{(5)}_1 - \lambda^{(2)}_{3}).
\end{split}
\end{align}
We now argue that it is not possible to have all six $\Gamma$-function to have zero or negative argument simultaneously. Precisely speaking, it is not possible for
\begin{align}
    \Gamma (1+ \lambda_2^{(3)} - \lambda_5^{(1)} ), \ \Gamma(2+\lambda^{(5)}_1 - \lambda^{(2)}_{3})
\end{align}
to have non-positive argument simultaneously. Since
\begin{align}
    \lambda^{(2)}_3 \leq \lambda^{(3)}_3 + 1 \leq \lambda^{(4)}_3+2=2 \implies 2+\lambda^{(5)}_1 - \lambda^{(2)}_3 \geq \lambda^{(5)}_1 \geq 0.
\end{align}
The bound is saturated when $\lambda^{(2)}_3=2$, which will require $\lambda^{(3)}_3=1$. 
On the other hand, 
\begin{align}
    \lambda^{(1)}_5 \leq \lambda^{(2)}_5 + 1 = 1 \implies 1 + \lambda^{(3)}_2 - \lambda^{(1)}_5 \geq \lambda^{(3)}_2 \geq 0.
\end{align}
The inequality is saturated when $\lambda^{(3)}_2 = 0$. The structure of the Young diagram will restrict $\lambda^{(3)}_3 \leq \lambda^{(3)}_2=0 \implies \lambda^{(3)}_3=0$. Hence it is not possible for the two $\Gamma$-functions to have zero argument simultaneously.

There are six $\Gamma$-functions in the denominator in \eqref{eq:analysis}:
\begin{align}
\begin{split}
    & \prod_{\omega=4}^6 \prod_{\beta+3 \leq \omega} \Gamma \left( n_\beta - n_{\beta+2} - 1 + \lambda^{(\beta+2)}_{\omega-\beta-2} - \lambda_{\omega-\beta}^{(\beta)} \right) \\
    & =  \Gamma (1+ \lambda_1^{(3)} - \lambda_3^{(1)} )  \Gamma (1+ \lambda_2^{(3)} - \lambda_4^{(1)} ) \Gamma (1+ \lambda_3^{(3)} - \lambda_5^{(1)} )   \\
    & \quad \times \Gamma (1+ \lambda_1^{(4)} - \lambda_3^{(2)} ) { \Gamma (1+ \lambda_2^{(4)} - \lambda_4^{(2)} ) } \Gamma (1+ \lambda_1^{(5)} - \lambda_3^{(3)} ) 
\end{split}
\end{align}
Notice that all potential poles coming from the $\Gamma$-function in the numerator will be canceled by the following arrangement:
\begin{align}
\begin{split}
    & \frac{\Gamma (1+ \lambda_1^{(3)} - \lambda_4^{(1)} )}{\Gamma (1+ \lambda_1^{(3)} - \lambda_3^{(1)} )} 
    \times \frac{\Gamma (1+ \lambda_2^{(3)} - \lambda_5^{(1)} )}{\Gamma (1+ \lambda_3^{(3)} - \lambda_5^{(1)} )} 
    \times \frac{\Gamma (1+ \lambda_1^{(4)} - \lambda_4^{(2)} )}{\Gamma (1+ \lambda_2^{(4)} - \lambda_4^{(2)} )} \\
    & \times \frac{\Gamma(2+\lambda^{(4)}_1 - \lambda^{(1)}_3 )}{\Gamma (1+ \lambda_1^{(4)} - \lambda_3^{(2)} )} 
    \times \frac{\Gamma (2+ \lambda_2^{(4)} - \lambda_4^{(1)} )}{\Gamma (1+ \lambda_2^{(3)} - \lambda_4^{(1)} )} 
    \times \frac{\Gamma(2+\lambda^{(5)}_1 - \lambda^{(2)}_{3})}{\Gamma (1+ \lambda_1^{(5)} - \lambda_3^{(3)} )}
\end{split}
\end{align}
It's easy to check that the argument of $\Gamma$-functions in the numerator is always greater or equal to the argument of the $\Gamma$-function in the denominator. Thus the pseudo measure is always finite in the $\kappa \to -\frac{1}{3}$ limit. 

\paragraph{}
We notice that, based on the two examples, not all $\Gamma$-function in the numerator in \eqref{eq:analysis} can have non-positive argument. The number of $\Gamma$-functions generating potential poles in the $\kappa \to -\frac{r-1}{k+1}$ limit is limited and will be canceled by the $\Gamma$-function coming from the denominator in the first line of \eqref{eq:analysis}. The pseudo-measure of all instanton configurations stay finite and the Jack polynomial does not have pole at $\kappa \to - \frac{r-1}{k+1}$ when the partition $\bn$ is $(k,r)$-admissible.
This is easy to check when a partition $\bn$ is given. The proof for general situation is rather subtle. It will be nice to have rigorous proof for general case from the gauge theory.

\subsubsection{Gauge theory perspective: further Higgsing}

We have seen that the specialization of the parameter $\kappa$~\eqref{kappa_admissible} leads to the admissible condition.
On the other hand, this parameter $\kappa$ can be written in terms of the $\Omega$-background parameters as mentioned in~\eqref{nu_epsilon}.
Hence, we may write the condition on $\kappa$ as follows,
\begin{align}
    q_1^{r-1} q_4^{-(k+1)} = 1
    \, ,
\end{align}
where we parametrize the $\Omega$-background parameters as $q_i = e^{\ve_i}$ for $i = 1,\dots,4$. 
This expression implies that the admissible condition may be interpreted as the Higgsing process in $\mathbb{C}_1 \times \mathbb{C}_4$ from the point of view of gauge origami.
A similar situation has been studied in \cite{Kimura:2022spi}%
\footnote{%
We remark that the NS limit is taken in our case, $q_2 \to 1$ ($\ve_2 \to 0$).
}
that points out that such a Higgsing condition is interpreted as the resonance condition~\cite{Feigin:2010qea,Feigin:2013JA} in the context of quantum toroidal algebras.

\section{Discussion and Future Direction} \label{sec:discussion}

In this paper we have established relations between three objects: the surface operator of 4d gauge theory, the Jack polynomials, and fractional quantum Hall states. In particular, the main result is to realize the fractional quantum Hall states as the instanton partition function of four dimensional $\CalN=2^*$ in the presence of full-type surface defect (up to an overall Gaussian factor) in the following simultaneous limits:

\begin{itemize}
    \item Using the $qq$-character we are able to identify the instanton partition function of four dimensional $\CalN=2^*$ supersymmetric gauge theory in the presence of surface defects as the eigenfunction of the $N$-body elliptic Calogero-Moser system in the Nekrasov-Shatashivili limit $\ve_2 \to 0$ (\ref{cond:NS}).
    \item The trigonometric limit $\kq = e^{2\pi i \tau} \to 0$ (\ref{condi:Tri}) of the elliptic Calogero-Moser system, which translates to the bulk decoupling limit in the gauge theory, simplifies the gauge theory partition function to the surface contribution. The defect instanton partition function is then proven to be the eigenfunction of the Laplace-Beltrami operator.
    \item With proper Higgsing condition (\ref{condi:Higgs}) imposed on the Coulomb moduli parameters, the defect supersymmetric partition function is identified with the Jack polynomial, with the defining partition given by the quantization condition. 
    On the side of four dimensional gauge theory, the presence of both the orbifolding and Higgsing can be understood as two different types of the co-dimensional two surface defects are introduced simultaneously. 
    \item We also explored the reconstruction of Laughlin and Moore-Read states from the defect instanton partition function (\ref{condi:filling}). It is well known that Laughlin and Moore-Read states serve as models for the study of both abelian and non-abelian quantum Hall effect (up to an overall Gaussian factor) with the filling fraction given by $\nu=\frac{\ve_3}{\ve_1}$. 
\end{itemize}


The translation from the defect partition function (with bulk decoupled) to the FQH state wavefunction requires a Gaussian factor shared by all FQH states:
\begin{align}
    \exp \left( - \frac{eB}{4\hbar} \sum_{\alpha=1}^N |z_\alpha|^2 \right)
\end{align}
here $B$ is the magnetic field, $e$ is the electron charge, $\hbar=\ve_1$ is the Planck constant. In section \eqref{sec:Quantization} we placed the trigonometric Calogero-Moser system on a circle with $z_\alpha = e^{2\pi i \rx_\alpha}$. The Gaussian factor is nothing but an overall constant since $|z_\alpha|=1$ for all $\alpha=1,\dots,N-1$. It would be nice to see in a more general case whether the Gaussian factor can be realized physically from the gauge theory side. 
 
We would like to note that our construction of the FQHE from 4d $\CalN=2^*$ theory has similarity but not equivalent to the construction in \cite{Vafa:2015euh}. In the later, the ADE $\CalN=(2,0)$ gauge theory in 6d lives on $S^3_{{\ve_2}/{\ve_1}} \times \BR \times \Sigma$. $S^3_{{\ve_2}/{\ve_1}}$ is a squashed 3-sphere. $\Sigma$ is a 2d Riemann sphere known as the Gaiotto curve. In particular, the FQHE filling fraction is identified as $\nu-2\frac{\ve_2}{\ve_1}$ in \cite{Vafa:2015euh}. 

The $qq$-character observables \eqref{def:A0qq} are known to have analytic property on its argument. We would like to know if one can use the analytic property of $qq$-character to prove the admissible condition of the Jack polynomial: a Jack polynomial $J^{\frac{1}{\kappa}}_\bn$ is regular at $\kappa = - \frac{r-1}{k+1}$ when the partition $\bn$ is $(k,r)$-admissible. 
The admissible condition has been proven using the clustering properties of the Jack polynomial \cite{desrosiers2015jack}. 
The hardship lies on the fact that the analytic property of $qq$-character is associated to its argument $x$, which in the context of gauge origami is the moduli parameter on the auxiliary space $\BC_{34}^2$. On the other hand the Jack parameter $\kappa$ that the admissible condition addresses is associated to the adjoint mass $m$ and $\Omega$-deformation parameter $\ve_1$ in the gauge theory. Both are free parameters in the gauge theory. It will be very helpful if gauge theory can provide a proof of the admissible condition of the Jack polynomial: 

It is known that the supersymmetric gauge theory instanton partition function has five and six dimensional extension. Using the same strategy of the $qq$-character one should recover the Macdonald polynomial and its elliptic uplift. Furthermore, if a proof of the admissible condition for Jack polynomial can be shown from the corresponding 4 dimensional gauge theory, it is possible that an admissible condition would be applied to both Macdonald polynomial and its elliptic uplift using the five and six dimensional gauge theories. 

\section*{Acknowledgement}
The work of TK was supported by ``Investissements d'Avenir'' program, Project ISITE-BFC (No.~ANR-15-IDEX-0003), EIPHI Graduate School (No. ANR-17-EURE-0002), and Bourgogne-Franche-Comté region.
The work of NL was supported by IBS-R003-D1. NL would like to thank N. Nekrasov and A. Grekov for discussion. 

\newpage

\appendix

\section{Special Functions}\label{sec:functions}

\subsection{Random Partition}
A partition is defined as a way of expressing a non-negative integer $n$ as summation over other non-negative integers. Each partition can be labeled by a Young diagram $\lambda=(\lambda_1,\lambda_2,\dots,\lambda_{\ell(\lambda)})$ with $\lambda_i\in\mathbb{N}$ such that 
\begin{equation}
n=|\lambda|=\sum_{i=1}^{\ell(\lambda)}\lambda_i.
\end{equation} 
We define the generating function of such a partition as
\begin{subequations}\label{phi}
\begin{align}
&\sum_{\lambda}\mathfrak{q}^{|\lambda|}=\frac{1}{(\mathfrak{q};\mathfrak{q})_\infty},\quad(\mathfrak{q};\mathfrak{q})_\infty=\prod_{n=1}^\infty\left(1-\mathfrak{q}^n\right); \\
&\sum_{\lambda}t^{\ell(\lambda)}\mathfrak{q}^{|\lambda|}=\frac{1}{(\mathfrak{q}t;\mathfrak{q})_\infty};\quad (\mathfrak{q}t;\mathfrak{q})_\infty=\prod_{n=1}^\infty(1-t\mathfrak{q}^n).
\end{align}
\end{subequations}
The $\mathfrak{q}$-shifted factorial (the $\mathfrak{q}$-Pochhammer symbol) is defined as
\begin{equation}\label{q-Pochhammer}
(z;q)_n=\prod_{m=0}^{n-1}(1-zq^m) .
\end{equation}

\subsection{Elliptic Functions}
\paragraph{}
Here we fix our notation for the elliptic functions. The so-called Dedekind eta function is denoted as
\begin{equation}\label{eta}
\eta(\tau)=e^{\frac{\pi i\tau}{12}}(\mathfrak{q};\mathfrak{q})_\infty .
\end{equation}
The first Jacobi $\theta$ function is denoted as:
\begin{equation}\label{theta}
\theta_{11}(z;\tau)=ie^{\frac{\pi i\tau}{4}}z^{\frac{1}{2}}(\mathfrak{q};\mathfrak{q})_\infty (\mathfrak{q}z;\mathfrak{q})_\infty(z^{-1};\mathfrak{q})_\infty,
\end{equation} 
whose series expansion 
\begin{equation}\label{theta2}
\theta_{11}(z;\tau)=i\sum_{r\in\mathbb{Z}+\frac{1}{2}}(-1)^{r-\frac{1}{2}}z^re^{\pi i\tau r^2}=i\sum_{r\in\mathbb{Z}+\frac{1}{2}}(-1)^{r-\frac{1}{2}}e^{rx}e^{\pi i\tau r^2},
\end{equation}
implies that it obeys the heat equation
\begin{equation}
\frac{1}{\pi i}\frac{\partial}{\partial\tau}\theta_{11}(z;\tau)=(z\partial_z)^2\theta_{11}(z;\tau).
\end{equation}
We also use another convention for the theta function,
\begin{align}
    \theta(z;\mathfrak{q}) = (z;\mathfrak{q})_\infty (\mathfrak{q}/z;\mathfrak{q})
    \, ,
    \label{theta3}
\end{align}
which has a different normalization from the previous definition \eqref{theta}.
The Weierstrass $\wp$-function
\begin{align}\label{Def:p-function}
\wp(z) = \frac{1}{z^2}+\sum_{p, q \ge 0} \left\{\frac{1}{(z+p+q\tau)^2}-\frac{1}{(p+q\tau)^2}\right\},
\end{align} 
is related to theta and eta functions by
\begin{equation}
\wp(z;\tau)=-(z\partial_z)^2\log\theta_{11}(z;\tau)+\frac{1}{\pi i}\partial_\tau\log\eta(\tau).
\end{equation}
We define the elliptic $\Gamma$-function,
\begin{align}
    \Gamma(z;p,q) = \prod_{0 \le n,m \le \infty} \frac{1 - z^{-1} p^{n+1} q^{m+1}}{1 - z p^n q^m}
    \label{eq:ell_gamma}
\end{align}
obeying the functional relation,
\begin{align}
    \frac{\Gamma(qz;p,q)}{\Gamma(z;p,q)} = \theta(z;p)
    \, , \qquad
    \frac{\Gamma(pz;p,q)}{\Gamma(z;p,q)} = \theta(z;q)
    \, .
\end{align}

\subsection{Higher rank Theta function}
Let us define
\begin{equation}
\Theta_{A_{N-1}}(\vec{z};\tau)=\eta(\tau)^{N}\prod_{\alpha>\beta}\frac{\theta_{11}(z_\alpha/z_\beta;\tau)}{\eta(\tau)}
\end{equation}
as the rank $N-1$ theta function, which also satisfies the heat equation \cite{Kac}
\begin{equation}\label{heat}
N\frac{\partial}{\partial\tau}\Theta_{A_{N-1}}(\vec{z};\tau)=\pi i\Delta_{\vec{z}}\Theta_{A_{N-1}}(\vec{z};\tau),
\end{equation}
with the $N$-variable Laplacian:
\begin{align}
    \Delta_{\vec{z}}=\sum_{\omega=0}^{N-1}(z_\omega\partial_{z_\omega})^2.
\end{align}

\subsection{Orbifolded Partition}
\paragraph{}
For the purpose in the main text, we consider the orbifolded coupling 
\begin{align}
    \mathfrak{q}=\prod_{\omega=0}^{N-1}\mathfrak{q}_\omega;\quad \mathfrak{q}_{\omega+N}=\mathfrak{q}_\omega,
\end{align}
and
\begin{align}
    \mathfrak{q}_\omega=\frac{z_\omega}{z_{\omega-1}};\quad z_{\omega+N}=\mathfrak{q}z_{\omega}.
\end{align}
We also consider the orbifolded version of the generating function of partitions $(\mathfrak{q};\mathfrak{q})_\infty^{-1}$ in \eqref{phi}. Given a finite partition $\lambda=(\lambda_1,\dots,\lambda_{\ell(\lambda)})$, we define
\begin{equation}
\mathbb{Q}^{\lambda}_\omega
=\prod_{j=1}^{\lambda_1}\mathfrak{q}_{\omega+1-j}^{\lambda_j^t}
=\prod_{i=1}^{\ell(\lambda)}\frac{z_\omega}{z_{\omega-\lambda_{i}}},
\end{equation}
where we used the relation \eqref{def:z}.
The summation over all possible partition is given by
\begin{equation}\label{Q-form}
\mathbb{Q}_\omega=\sum_{\lambda}\mathbb{Q}^{\lambda}_\omega=\sum_{\lambda}\prod_{i=1}^{\ell(\lambda)}\left(\frac{z_\omega}{z_{\omega-\lambda_{i}}}\right)=\sum_{l_0,\dots,l_{N-1},l\geq0}\prod_{\alpha=1}^{N-1}\left(\frac{z_\omega}{z_{\alpha}}\right)^{l_\alpha}\mathfrak{q}^l.
\end{equation}
The function $\mathbb{Q}(\vec{z};\tau)$ is the orbifolded version of the generating function of partitions \eqref{phi}, 
\begin{align}\label{Q}
\mathbb{Q}
&=\prod_{\omega=0}^{N-1}\mathbb{Q}_\omega(\vec{z};\tau) \nonumber\\
&=\prod_{N-1 \geq \alpha>\beta \geq 0}\frac{1}{(\frac{z_\alpha}{z_\beta};\mathfrak{q})_\infty(\mathfrak{q}\frac{z_\beta}{z_\alpha};\mathfrak{q})_\infty}\prod_{\alpha=0}^{N-1}\frac{1}{(\mathfrak{q};\mathfrak{q})_\infty} \nonumber\\
&=\prod_{N-1\geq\alpha>\beta\geq0}\frac{\mathfrak{q}^{1/12}\eta(\tau)\sqrt{z_\alpha/z_\beta}}{\theta_{11}(z_\alpha/z_\beta;\tau)}\times\left[\frac{\mathfrak{q}^{1/24}}{\eta(\tau)}\right]^N \nonumber\\
&=\left[\eta(\tau)^{-N}\prod_{N-1\geq\alpha>\beta\geq0}\frac{\eta(\tau)}{\theta_{11}(z_\alpha/z_\beta;\tau)}\right]\frac{\mathfrak{q}^{N^2/24}}{\vec{z}^{\vec{\rho}}} \nonumber\\
&=\frac{1}{\Theta_{A_{N-1}}(\vec{z};\tau)}\frac{\mathfrak{q}^{N^2/24}}{\vec{z}^{\vec{\rho}}},
\end{align}
where $\vec{\rho}$ is the Weyl vector of $SU(N)$ Lie group, whose entries are given as
\begin{equation}
\vec{\rho}=(\rho_0,\dots,\rho_{N-1});\quad\rho_\omega=\omega-\frac{N-1}{2};\quad |\vec{\rho}|^2=\sum_{\omega=0}^{N-1}\rho_\omega^2=\frac{N(N^2-1)}{12};\quad\vec{z}^{\vec{\rho}}=\prod_{\omega=0}^{N-1}z_\omega^{\rho_\omega}.
\end{equation}
Using eq.~\eqref{heat}, it is easy to prove that the $\mathbb{Q}$-function satisfies
\begin{equation}\label{Heat eq for Q}
0=\sum_{\omega}\nabla^{\mathfrak{q}}_\omega\log\mathbb{Q}-\frac{1}{2}\Delta_{\vec{z}}\log\mathbb{Q}+\frac{1}{2}\sum_\omega(\nabla^z_\omega\log\mathbb{Q})^2,
\end{equation}
with 
\begin{equation}
\sum_\omega\nabla_\omega^\mathfrak{q}=N\nabla^\mathfrak{q}+\vec{\rho}\cdot{\nabla}^{\vec{z}} .
\end{equation}





\newpage
\bibliographystyle{utphys}
\bibliography{QHEJack}

\end{document}